\documentclass{sig-alternate-10pt}

\usepackage{subfigure}
\usepackage{multirow}
\usepackage{algorithm}
\usepackage{algorithmicx}
\usepackage{algpseudocode}
\usepackage{array}
\usepackage{url} 
\usepackage{xcolor}
\usepackage{graphicx}
\usepackage{flushend}

\newcommand{\comment}[1]{}
\newcommand{\specialcell}[2][c]{%
  \begin{tabular}[#1]{@{}c@{}}#2\end{tabular}}

\begin{document}

\title{Improving Energy Efficiency of MPTCP for Mobile Devices}

\numberofauthors{3}
\author{
\alignauthor Yeon-sup Lim\\
       \affaddr{School of Computer Science} \\
       \affaddr{University of Massachusetts}\\
       \affaddr{Amherst, MA, USA}\\
       \email{ylim@cs.umass.edu}
\alignauthor Yung-Chih Chen\\
       \affaddr{School of Computer Science} \\
       \affaddr{University of Massachusetts}\\
       \affaddr{Amherst, MA, USA}\\
       \email{yungchih@cs.umass.edu}
\alignauthor Erich M. Nahum \\
       \affaddr{IBM Thomas J. Watson Research Center}\\
       \affaddr{Yorktown Heights, NY, USA}\\
       \email{nahum@us.ibm.com}
\and
\alignauthor Don Towsley\\
       \affaddr{School of Computer Science} \\
       \affaddr{University of Massachusetts}\\
       \affaddr{Amherst, MA, USA}\\
       \email{towsley@cs.umass.edu}
\alignauthor Richard J. Gibbens \\
       \affaddr{Computer Laboratory} \\
       \affaddr{University of Cambridge}\\
       \affaddr{Cambridge, UK}\\
       \email{richard.gibbens@cl.cam.ac.uk}
}

\maketitle

\begin{abstract}
Multi-Path TCP (MPTCP) is a new transport protocol that enables systems to exploit available paths through multiple network interfaces.
MPTCP is particularly useful for mobile devices, which usually have multiple wireless interfaces.
However, these devices have limited power capacity and thus judicious use of these interfaces is required.
In this work, we develop a model for MPTCP energy consumption derived from experimental measurements using MPTCP on a mobile device with both cellular and WiFi interfaces.
Using our energy model, we identify an operating region where there is scope to improve power efficiency compared to both standard TCP and MPTCP.
We design and implement an improved energy-efficient MPTCP, called eMPTCP. 
We evaluate eMPTCP on a mobile device across several scenarios, including varying bandwidth, background traffic, and user mobility.
Our results show that eMPTCP can reduce the power consumption by up to 15\% compared with MPTCP, while preserving the availability and robustness benefits of MPTCP. 
Furthermore, we show that when compared with TCP over WiFi, which is more energy efficient than TCP over LTE, eMPTCP obtains significantly better performance with relatively little additional energy overhead.
\end{abstract}
\let\thefootnote\relax\footnote{This paper is an extended version of a paper that has been accepted to ACM AllThingsCellular'14 \cite{ourcellnet}.}



\section{Introduction}
Multi-Path TCP (MPTCP) is a new standardized transport protocol
that simultaneously enables end hosts
to take advantage of multiple network interfaces and utilize path diversity in the network \cite{rfc6824, Raiciu2012}.
MPTCP can achieve both higher throughput and greater robustness and
availability than standard single-path TCP, all while maintaining
compatibility with existing applications. One natural fit for MPTCP is use on mobile devices, such as smartphones,
which typically include both cellular and WiFi interfaces.

Applying MPTCP to mobile devices introduces a new concern, namely,
the additional energy consumption from operating multiple network interfaces.
Mobile devices are frequently constrained by the amount of power available
in their batteries.
Processing power available on-chip continues to grow exponentially, however
battery storage increases slowly by comparison.
Thus, power consumption is an important area of research, particularly in
mobile devices such as smartphones.
In order to utilize MPTCP on mobile devices with limited energy resources,
it is important to understand the power consumption behavior of MPTCP
to ensure that it is practical.

In this work, we shed light on the energy behavior of MPTCP on smartphones.
We seek to address the following questions:

\begin{itemize}
\item How much energy does MPTCP consume, compared to single-path TCP over WiFi or LTE?  

\item Are there environments or scenarios where MPTCP is more energy efficient than single-path TCP?
If so, how can we recognize them and take advantage of them?

\item Can we improve MPTCP's energy efficiency?
\end{itemize}

In this work, we examine MPTCP energy consumption behavior via a combination of measurement, modeling, and experimentation.
We measure power consumption across a range of scenarios, by varying download/upload size and available path bandwidth.
We use these measurements together with a regression approach from \cite{Huang2012close}, and determine conditions under which
MPTCP is more energy-efficient than either standard TCP or MPTCP.
Informed by this model, we design and implement an energy-aware eMPTCP which we have implemented on an Android platform, the Samsung Galaxy S3.

Our paper makes the following contributions:

\begin{itemize}

\item We develop an energy model for MPTCP power consumption derived from experimental measurements taken using MPTCP on a mobile device.
Our results show that the model accurately predicts the measured energy consumption of MPTCP with an error less than 17\%.

\item Using our model, we illustrate the tradeoffs between network performance and energy consumption.
In most environments, MPTCP does not improve energy efficiency compared to single-path TCP over WiFi.
The high cost of cellular tail energy makes power saving difficult to achieve in MPTCP. 
However, our model does reveal an operating region where MPTCP is more energy efficient than standard TCP, depending on the available
path throughputs and transfer size.

\item We design, implement, and evaluate eMPTCP, an improved energy efficient MPTCP, on our Android mobile device. 
Parameterized through our model, eMPTCP dynamically monitors path characteristics and chooses paths based on energy efficiency.
We evaluate eMPTCP under multiple scenarios, considering path quality, dynamic bandwidth changes, radio interference, and mobility.
We show that eMPTCP can reduce energy consumption by up to 15\% compared with MPTCP while still preserving 
the availability and robustness benefits of multiple paths, at the cost of slightly longer download times.
\end{itemize}

Previous approaches in this area have either been simulation-only \cite{Pluntke2011} or have looked at much more restricted operating
environments \cite{Paasch2012, Raiciu2011:Opportunistic}.  
Our work provides general insight and leverages that understanding to provide a new and improved energy-aware MPTCP.

The remainder of this paper is organized as follows:
Section \ref{sec:background} provides the background context for our work.
Section \ref{sec:microbench} presents throughput and energy measurements
we use to parameterize a single-path TCP energy model.
Section \ref{sec:model} describes and validates our energy model for MPTCP.
We introduce our energy-aware MPTCP in Section \ref{sec:eamptcp}.
Section \ref{sec:evaluation} provides our results.
After reviewing related work in Section~\ref{sec:related},
we conclude in Section~\ref{sec:conclusion}.

\section{Background}
\label{sec:background}

\subsection{Benefits of Multi-path TCP}

The benefits of leveraging MPTCP in mobile devices are three-fold:

\noindent $\bullet$ \emph{Bandwidth}---By utilizing the available bandwidth of each subflow, a MPTCP connection can achieve higher throughput than a standard TCP connection.

\noindent $\bullet$ \emph{Robustness}---Even though connectivity in one network can degrade or disappear through movement, MPTCP offers a seamless TCP connection by using paths (subflows) through another network.

\noindent $\bullet$ \emph{Compatibility}---The MPTCP layer is hidden from user applications by providing a standard TCP socket. Existing applications using TCP need not to be modified to support MPTCP.

\subsection{3G/4G State Machine}
The 3rd Generation Partnership Project (3GPP) standard defines a state machine for the 3G/4G communication interface, which describes the possible power states of each device connected to the network. To initiate a packet transmission, a 3G/4G communication interface has to switch from a low power to high power state so that packet can be sent or received. If there is no further packet transmission for a period of time, to save energy, the 3G/4G interface returns to a low power state. In the rest of this paper, the terms \textit{promotion} and \textit{tail} are used to refer to the transition periods from a low power state to a high power state and from a high power state to a low power state, respectively \mbox{\cite{Balasubramanian2009, Huang2012close}.}

\subsection{Operation Modes in MPTCP}
MPTCP has three modes of operation to control subflow usage. One of them is the backup mode, where MPTCP opens TCP subflows over all interfaces, but uses only a subset of them for packet transmission. If a user sets a particular interface to backup mode, MPTCP sends no traffic through the corresponding subflows unless all other subflows break. By setting up the mode of each interface, a user can manually decide path usage considering traffic pricing or battery life \cite{Paasch2012}. However, MPTCP does not have any automatic mechanism to control the mode of each subflow in terms of energy efficiency and performance. In this paper, we develop an energy efficient path usage controller while MPTCP operates in Full-MPTCP mode (the regular MPTCP operation in which all subflows are available for use).

\section{Single-Path TCP Energy Model} \label{sec:microbench}
In this section, we develop an energy consumption model of single-path TCP for our mobile device. The energy model differs between mobile devices. We leave the development of energy models for other mobile devices as future work.

\begin{figure}[t++]
  \centering
  \subfigure[Setup]{\includegraphics[scale=0.23]{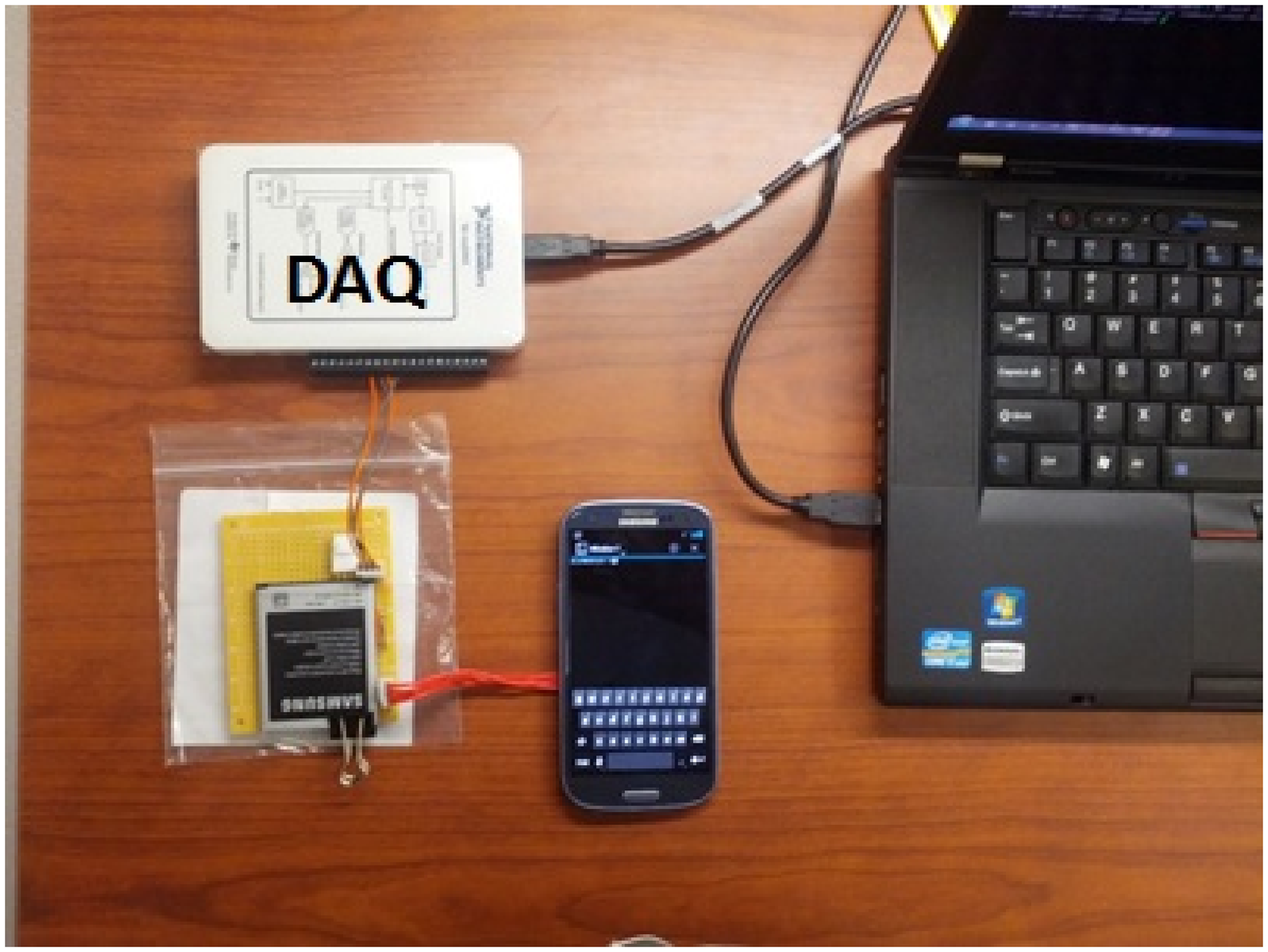}} \hfill
  \subfigure[Circuit]{\includegraphics[scale=0.23]{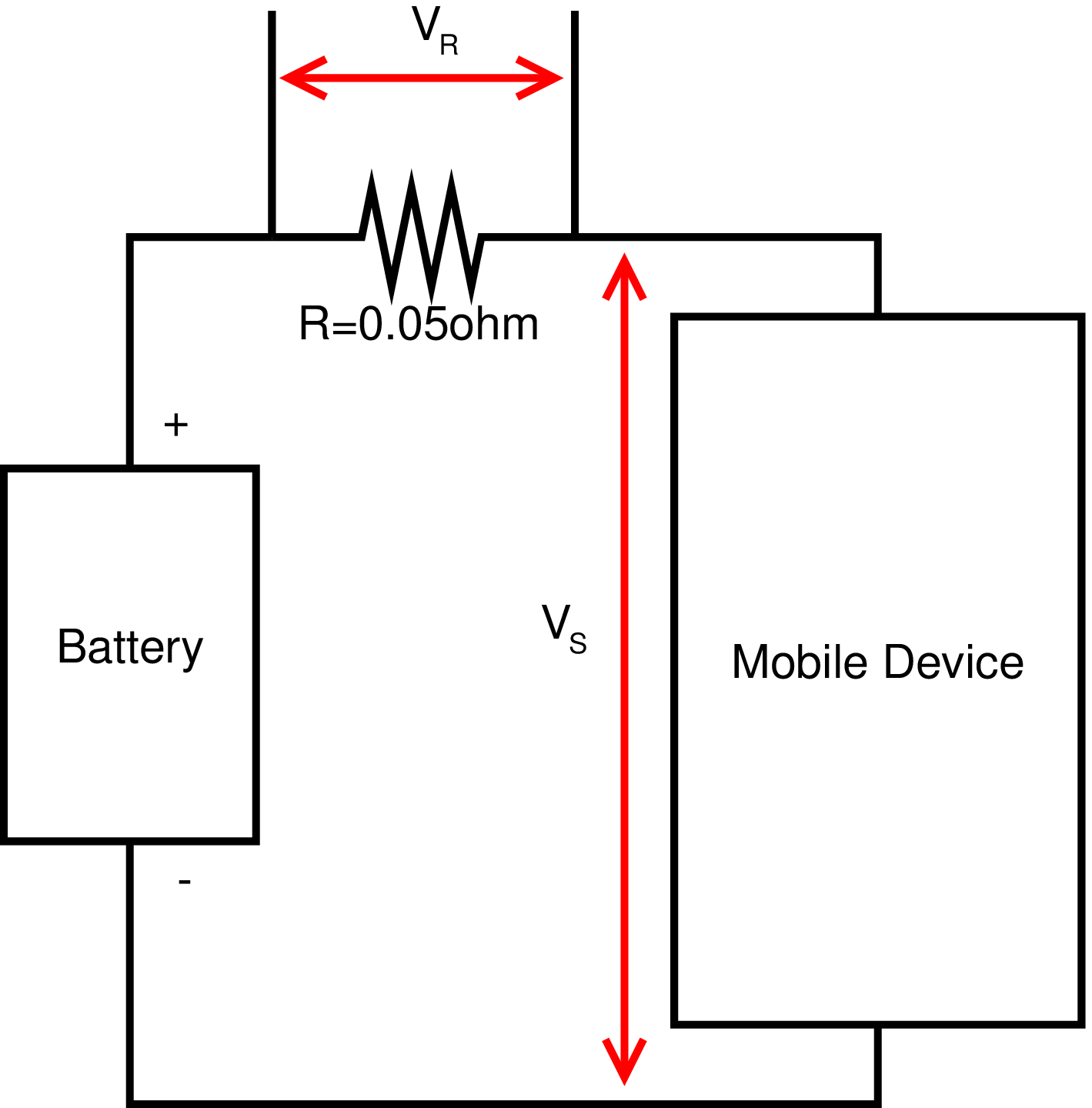}}
  \caption{Setup for Energy Profiling}
  \label{fig:benchmark}
\end{figure}

\subsection{Setup}
\label{sec:setup}
Our setup consists of a wired server (a desktop equipped with an Intel Quad Core I7-3770 CPU and 32 GB of memory) connected through a single Gigabit Ethernet interface to our campus network, a WiFi access point (IEEE 802.11g), and a mobile device (Samsung Galaxy S3 for AT\&T, SGH-I747).

The server is connected through a single Gigabit Ethernet interface to our campus network and runs Ubuntu Linux 12.04 with the MPTCP implementation \cite{Raiciu2012}. The mobile device is connected to the Internet using both 3G(HSDPA) or 4G(LTE) from AT\&T and WiFi. The Linux MPTCP kernel is ported into SGH-I747 running a customized Jellybean 4.1.2 platform using a 3.0.2. kernel~\cite{Lim}.

To measure the energy consumption of the mobile device, we build an electric circuit with an externally installed battery and a $R=0.05\Omega$ high precision resistor between the battery and the mobile device as shown in Figure \ref{fig:benchmark}. We use a National Instruments Data Acquisition system (NI-DAQ), with sampling rate of 100K samples/s, in order to measure the voltage supplied to the device ($V_S$) and the voltage drop across the resistor ($V_R$). The energy consumption at each measurement point ($P$) is calculated based on $R$, $V_R$ and $V_S$ as $P=\frac{V_R}{R} \times V_S$.
Energy consumption traces are collected during ten repeated uploads/downloads of files with size varying from 16KB to 16MB.
While profiling TCP energy consumption, we force the mobile device CPU to remain in performance mode so as to avoid energy consumption changes due to the CPU switching power modes.

\subsection{Profiling TCP Energy Consumption over a Single Interface}
\subsubsection{Promotion and Tail States}

Table \ref{tab:fixed_overhead} presents the average duration and energy consumption of the \textit{promotion} and \textit{tail} states measured from the collected traces.
The energy overheads due to the \textit{promotion} and \textit{tail} are constant for every single packet transfer starting from a low power state.
We refer to such overheads of each interface as the \textit{fixed} energy overhead.
WiFi has a comparatively short \textit{promotion} and \textit{tail} state, resulting in smaller fixed overheads compared with 3G (HSDPA) and 4G (LTE).

While the HSPDA and LTE \textit{tail} periods are similar, in the range of 12$\sim$16 seconds, the HSPDA \textit{promotion} period is around two seconds, roughly five times longer than that of LTE.
Since both the \textit{promotion} and \textit{tail} states of WiFi are much shorter, and the per-second energy cost is significantly less, the fixed overhead for WiFi is much lower than for the others.

\begin{table}[t!!!]
\scriptsize
\centering
\caption{Summary on Promotion and Tail States (Standard Deviations in Parentheses)}
\begin{tabular}{c | c | >{\centering}m{1.2cm} >{\centering}m{1.9cm}  >{\centering}m{1.2cm} }
\hline
\hline
\multicolumn{2}{c}{State} & Average Duration (sec) & Average Energy Consumption (J) & Fixed Energy Overhead (J) \tabularnewline
\hline
\multirow{2}{*}{HSDPA}  & Promotion & \specialcell{2.098\\($\pm$0.455)}   & \specialcell{1.463\\($\pm$ 0.306)}    & \multirow{2}{*}{11.337}  \tabularnewline
                        & Tail      & \specialcell{16.123\\($\pm$1.137)}  & \specialcell{9.873\\($\pm$1.057)}     & \tabularnewline
\hline
\multirow{2}{*}{LTE}    & Promotion & \specialcell{0.405\\($\pm$0.047)}   & \specialcell{0.311\\($\pm$0.041)}    & \multirow{2}{*}{2.908}  \tabularnewline
                        & Tail      & \specialcell{11.490\\($\pm$0.492)}  & \specialcell{2.597\\($\pm$0.275)}    & \tabularnewline
\hline
\multirow{2}{*}{WiFi}   & Promotion & \specialcell{0.095\\($\pm$0.029)}   & \specialcell{0.040\\($\pm$ 0.017)}    & \multirow{2}{*}{0.149}  \tabularnewline
                        & Tail      & \specialcell{0.295\\($\pm$0.152)} & \specialcell{0.109\\($\pm$ 0.080)}     & \tabularnewline
\hline
\end{tabular}
\label{tab:fixed_overhead}
\end{table}

\begin{table}[t!!!]
\scriptsize
\centering
\caption{Coefficients for Packet Transfer State}
\begin{tabular}{c  c >{\centering}m{1.3cm} >{\centering}m{2cm}  >{\centering}m{1.3cm} }
\hline
\hline
 \multicolumn{2}{c}{ \multirow{2}{*}{State} } & \multicolumn{3}{c}{Interface} \tabularnewline
 & & HSDPA & LTE & WiFi \tabularnewline
\hline
\multirow{2}{*}{Download}  & \multicolumn{1}{|c|}{$\alpha^d$} & 9.3440 & 10.0427    & 4.6750  \tabularnewline
                        & \multicolumn{1}{|c|}{$\beta^d$}      & -0.9286  & -0.8910     & -0.8179 \tabularnewline
\hline
\multirow{2}{*}{Upload}    & \multicolumn{1}{|c|}{$\alpha^u$} & 12.5294  & 13.3438 & 3.6135 \tabularnewline
                        & \multicolumn{1}{|c|}{$\beta^u$}     & -0.8524  & -0.8358 & -0.6617 \tabularnewline
\hline
\end{tabular}
\label{tab:coeff}
\end{table}

\begin{figure*}[ht!!!]
  \centering
  \subfigure[HSDPA]{ \includegraphics[scale=0.27]{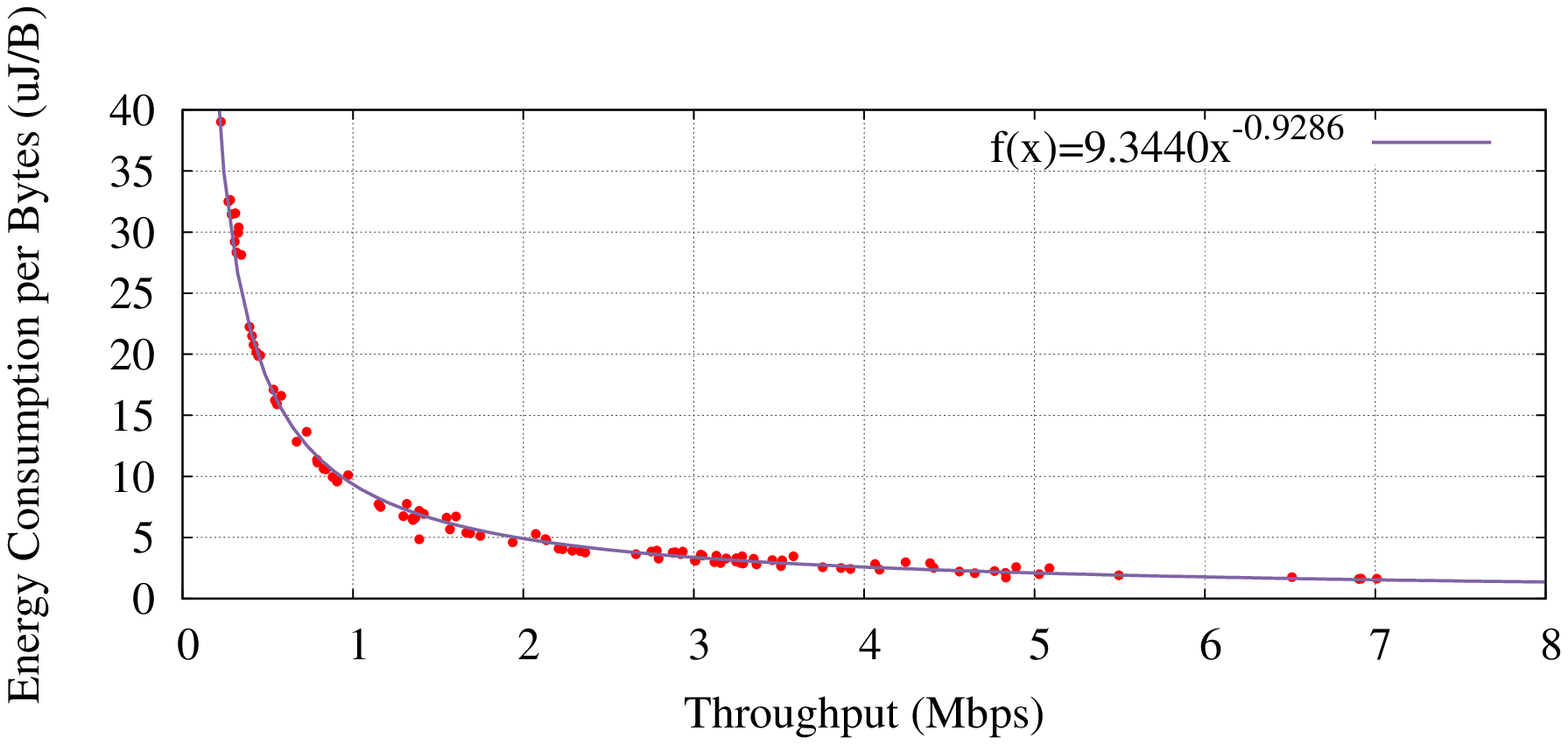}  }
  \subfigure[LTE]{ \includegraphics[scale=0.27]{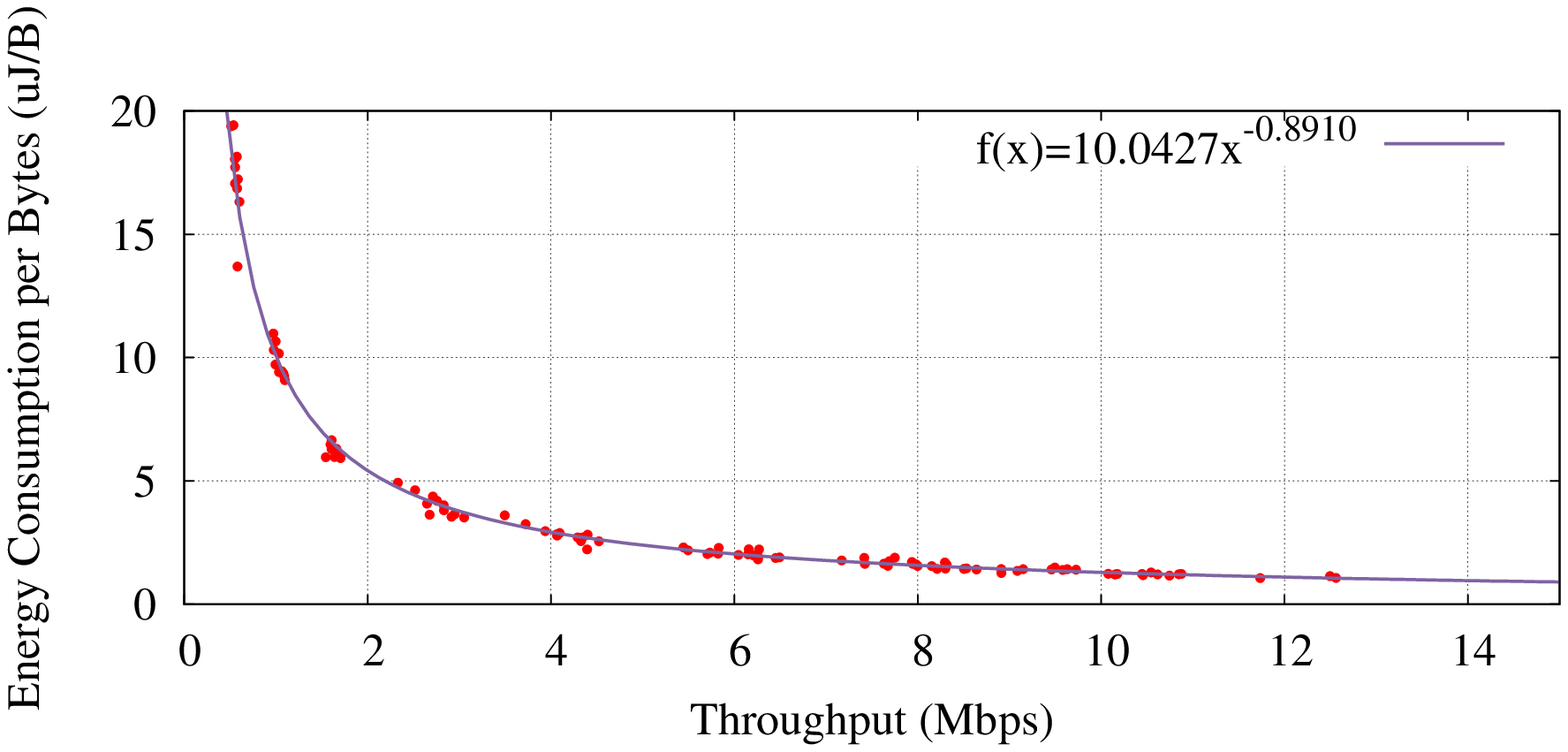}  }
  \subfigure[WiFi]{ \includegraphics[scale=0.27]{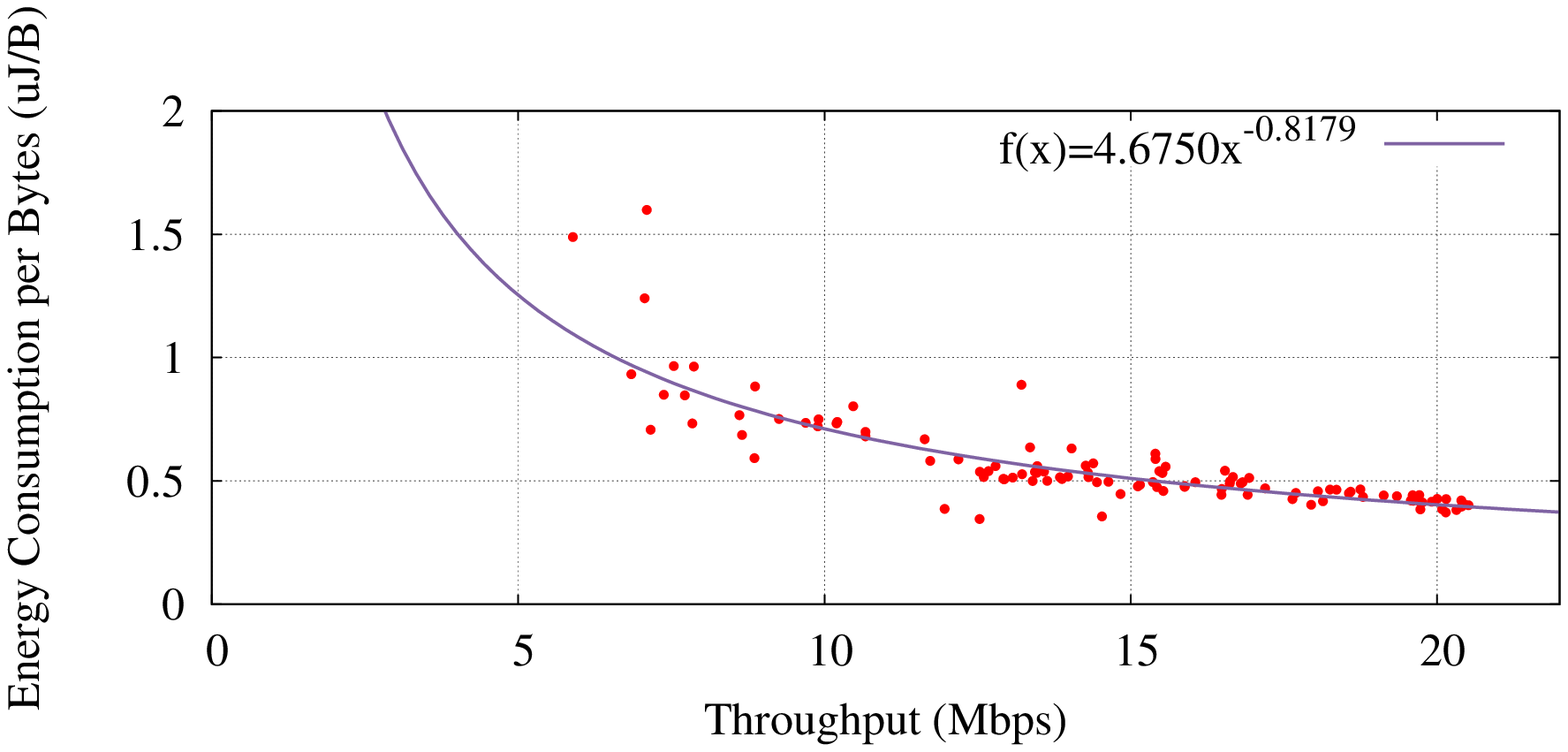}  }
  \caption{Energy Consumption per Byte during Packet Transfer - Downloads}
  \label{fig:coeff-download}
\end{figure*}

\begin{figure*}[ht!!!]
  \centering
  \subfigure[HSDPA]{ \includegraphics[scale=0.27]{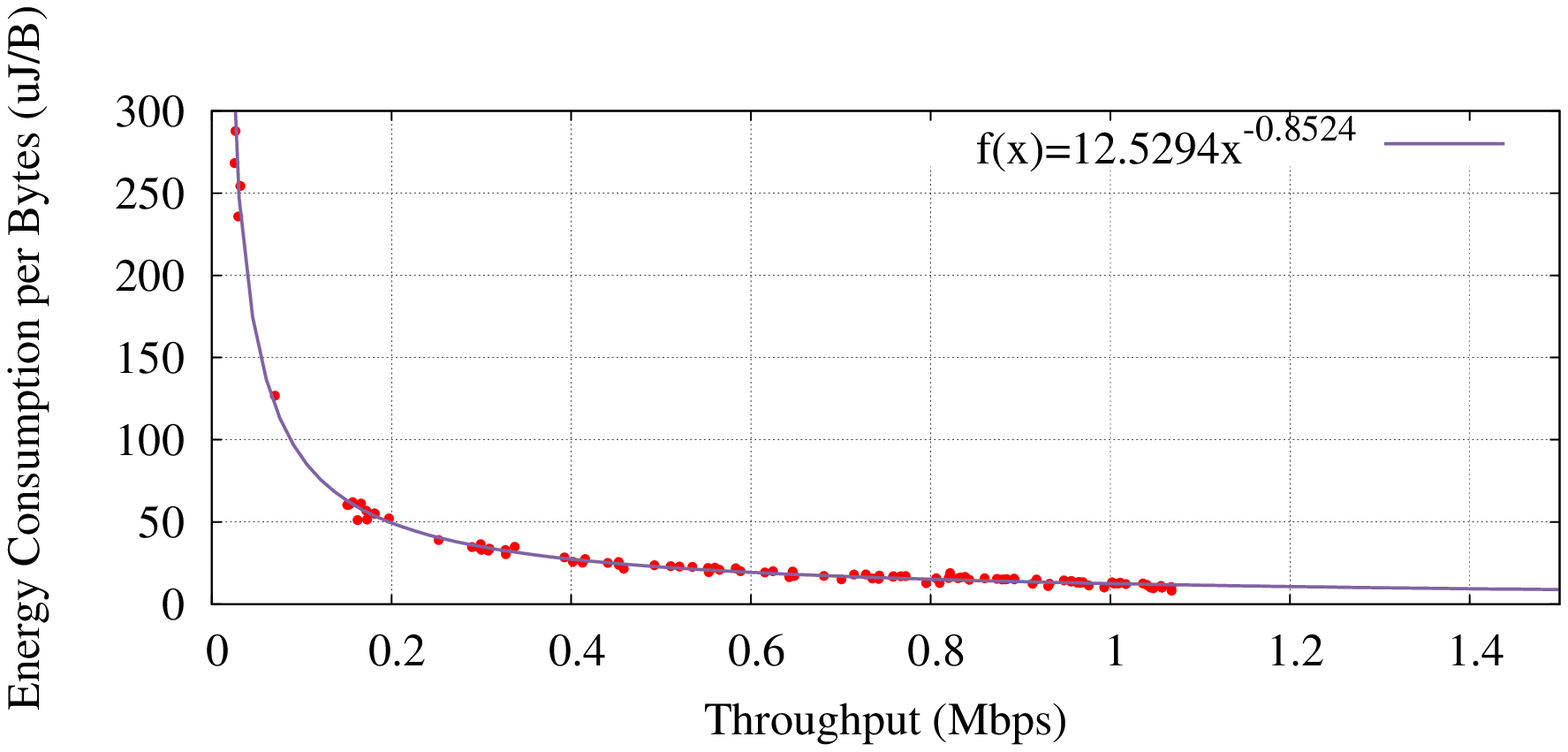}  }
  \subfigure[LTE]{ \includegraphics[scale=0.27]{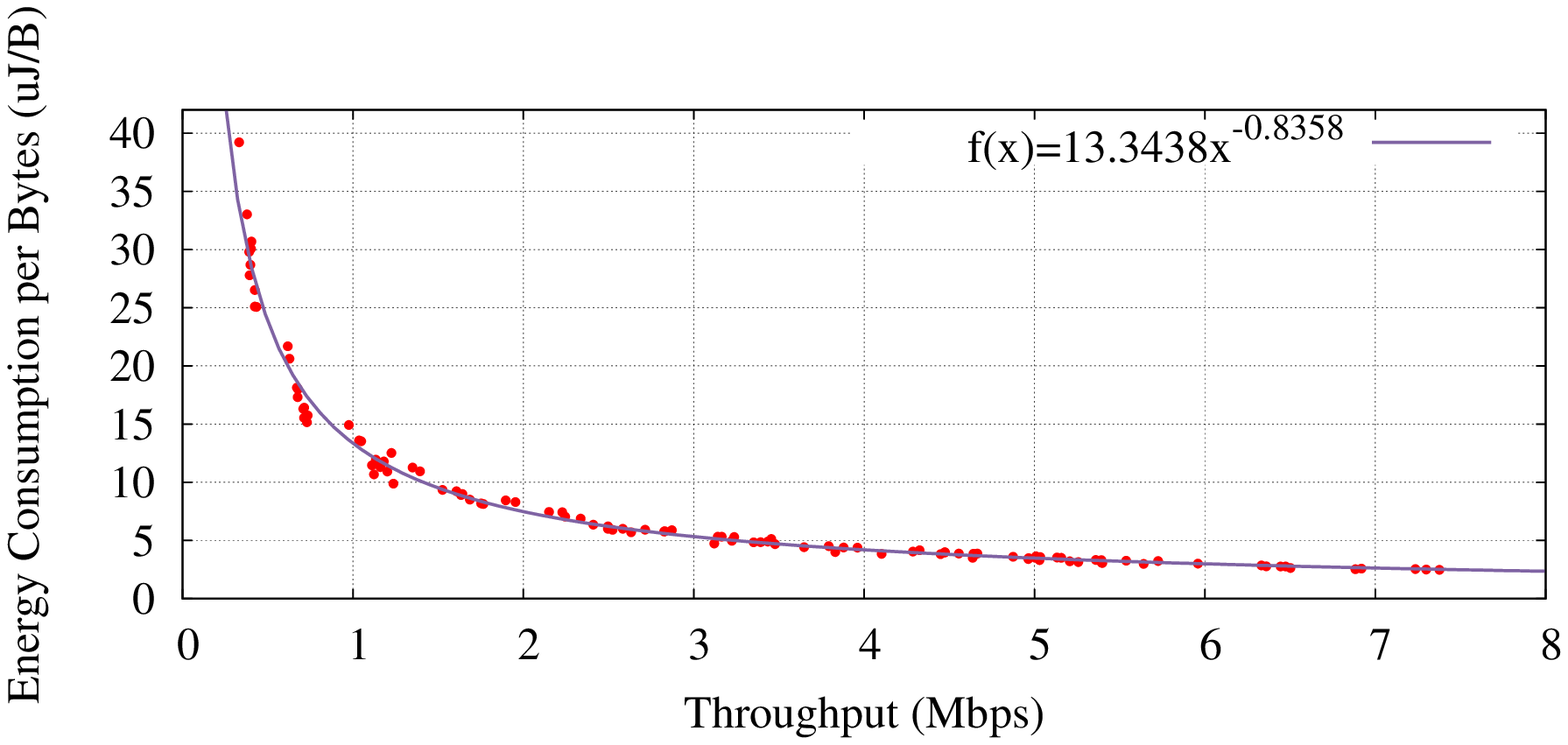}  }
  \subfigure[WiFi]{ \includegraphics[scale=0.27]{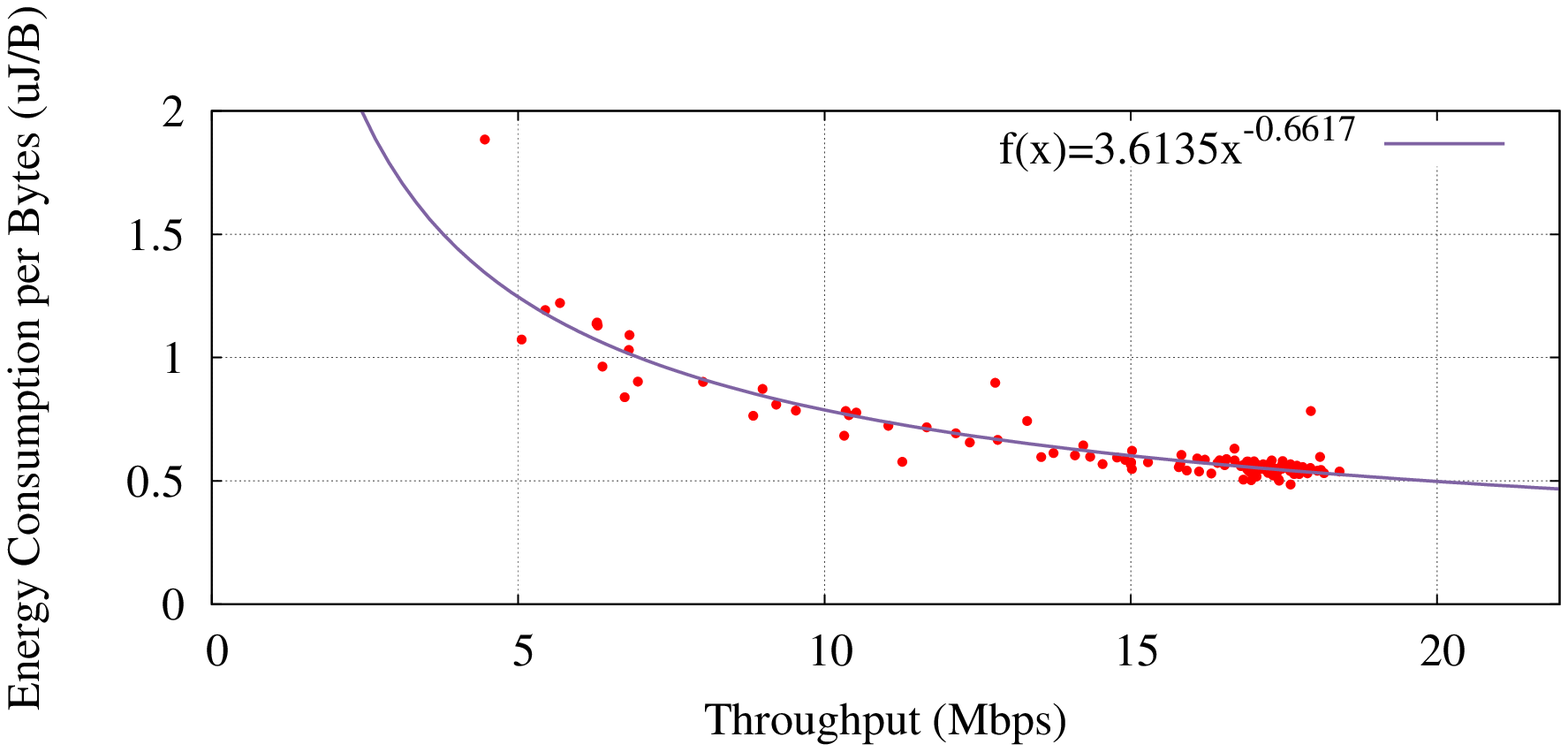}  }
  \caption{Energy Consumption per Byte during Packet Transfer - Uploads}
  \label{fig:coeff-upload}
\end{figure*}

\subsubsection{Packet Transfer State}
We use a simple energy model \cite{Balasubramanian2009, Huang2012close} with available bandwidth as the input parameter to model energy consumption during the packet transfer state. We assume that the energy consumption per transferred byte can be represented as a function of the upload/download TCP throughput. To estimate this function for each interface, we explore the energy consumption per transferred byte during packet transfer.

Figures \ref{fig:coeff-download} and \ref{fig:coeff-upload} illustrate the measured energy consumption per transferred byte as a function of obtained TCP throughputs. 
We observe that the regression model $P = \alpha \times B^{\beta}$ yields estimates of $P$ close to actual measured values as a function of $B$, the available throughput. Table \ref{tab:coeff} lists the estimates of $\alpha$ and $\beta$ for each setting. For example, given a download throughput over the LTE interface of $B_{L}$ (Mbps), the energy consumption for downloading each byte, $P_{L}$ $(\mu J/B)$, is defined as:
\begin{eqnarray*}
    P_{L} ( B_{L} ) &=& \alpha_{L} \times {B_{L}}^{\beta_{L}},
\end{eqnarray*}
where $\alpha_{L}=10.04$ and $\beta_{L}=-0.89$.
Here, we do not separately consider the energy consumption due to sending TCP ACK packets due to their small impact: since the size of the ACK packet is comparatively small, we consider the consumed energy for sending an ACK as part of the overall energy consumption for receiving data packets.

\begin{figure}[t!!!]
  \centering
  \includegraphics[scale=0.4]{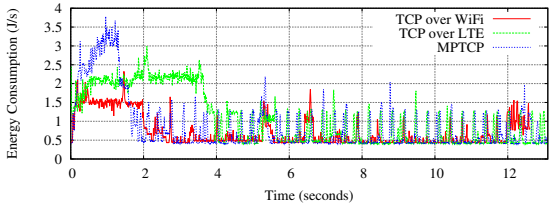}
  \caption{Sample Energy Trace - 4MB Download}
  \label{fig:energy-example}
\end{figure}

\section{MPTCP Energy Model} \label{sec:model}
Here we present our MPTCP energy model for the case that MPTCP simultaneously uses the LTE and WiFi interfaces based on the profiling results in the previous section. 

\subsection{Model}
Figure \ref{fig:energy-example} shows sample energy trace for TCP over each interface and MPTCP when the device downloads a 4MB file. We first observe that MPTCP consumes more energy during the first 2 seconds of data transfer. We also see that both TCP over LTE and MPTCP incur energy costs for the LTE \textit{tail} state (TCP over LTE starts its tail at around 4 seconds, and MPTCP starts it at around 2 seconds). Analogous to the energy model of standard TCP over each interface \cite{Huang2012close}, we model MPTCP energy consumption as a function of the available throughput of each interface.

Let $B_{L}$ and $B_{W}$ denote the throughputs of LTE and WiFi, respectively. Suppose that $S$ is the size of the file and $S_{W}$ and $S_{L}$ are the number of bytes transferred through WiFi and LTE, respectively; $S=S_{W} + S_{L}$. A simple estimate of MPTCP energy consumption is to sum the energy consumed by each interface over transferred packets. However, we observe that the MPTCP energy consumption is slightly less than this, possibly due to the shared use of energy. To consider such shared energy consumption, we assume that a device consumes a fraction $\gamma$ of the sum during the overlapped period for packet transfer. In the case of the fixed energy overhead for \textit{promotion} and \textit{tail}, we assume that the overhead is separately consumed for each interface. Let $\theta$ denote the ratio of the duration of the data transfer when both interfaces are simultaneously transferring packets. Given $S_{W}$ and $S_L$, we approximate $\theta$ as:

\begin{eqnarray*}
\small
    \theta = \dfrac{ \min{\left( S_{W}/B_{W}, S_{L}/B_{L} \right)} } { \max{\left( S_{W}/B_{W}, S_{L}/B_{L} \right)} },
\end{eqnarray*}

Based on our assumption, we estimate the MPTCP energy consumption during packet transfer, $E_{T}$, as:
\begin{eqnarray*}
\small
    E_{T} &=& \left(  P_{W} ( B_{W} ) \times S_{W} + P_{L} ( B_{L} ) \times S_{L} \right) \left( 1-\theta + \gamma \theta \right),
\end{eqnarray*}
where $B_W$ and $B_L$ are the available throughputs over WiFi and LTE and $S_W$ and $S_L$ are transferred bytes over WiFi and LTE, respectively.
The total MPTCP energy consumption, including the energy overheads associated with the \textit{promotion} and \textit{tail} states of each interface, are then represented as:
\begin{eqnarray*}
\small
    E_{M} &=& E_{T} + C_{W} + C_{L},
\end{eqnarray*}
where $C_{W}$ and $C_{L}$ are the fixed energy overheads for the \textit{promotion} and \textit{tail} states of the WiFi and LTE interfaces, respectively.


\begin{figure}[t!!!]
  \centering
  \subfigure[Downloads]{\includegraphics[scale=0.32]{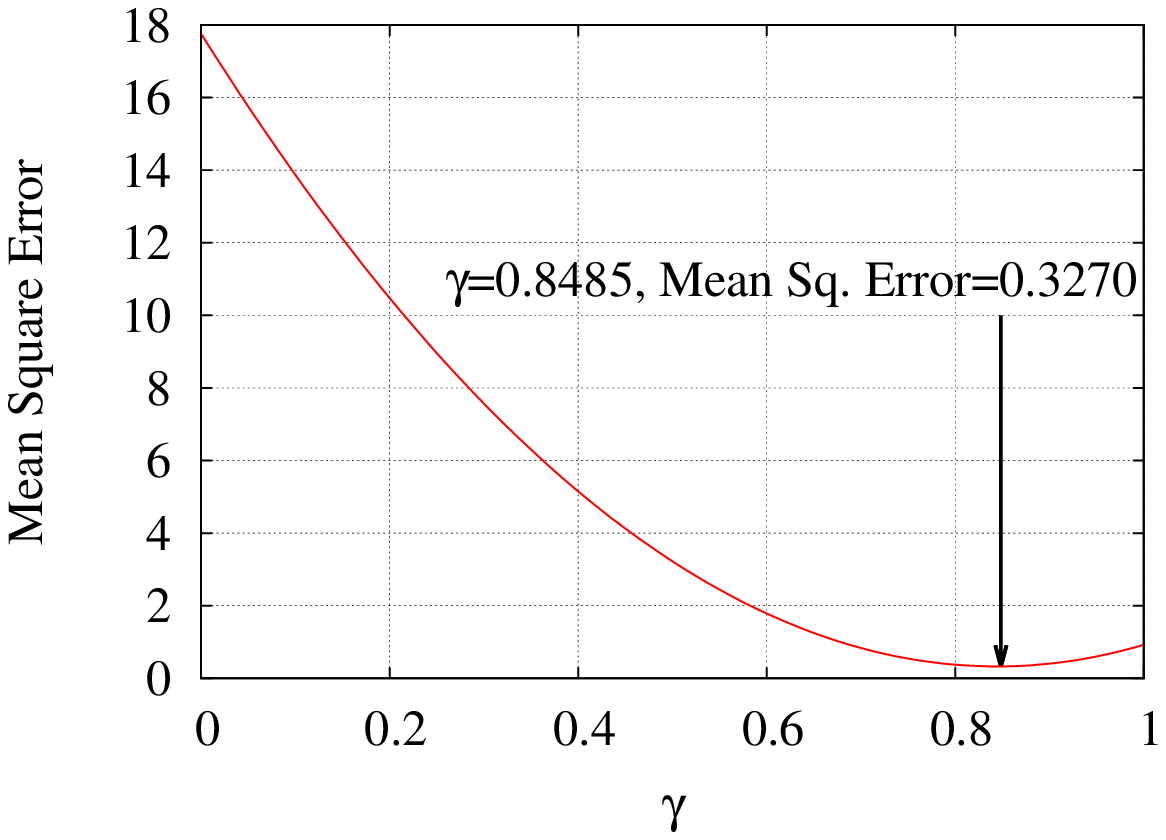}}
  \subfigure[Uploads]{\includegraphics[scale=0.32]{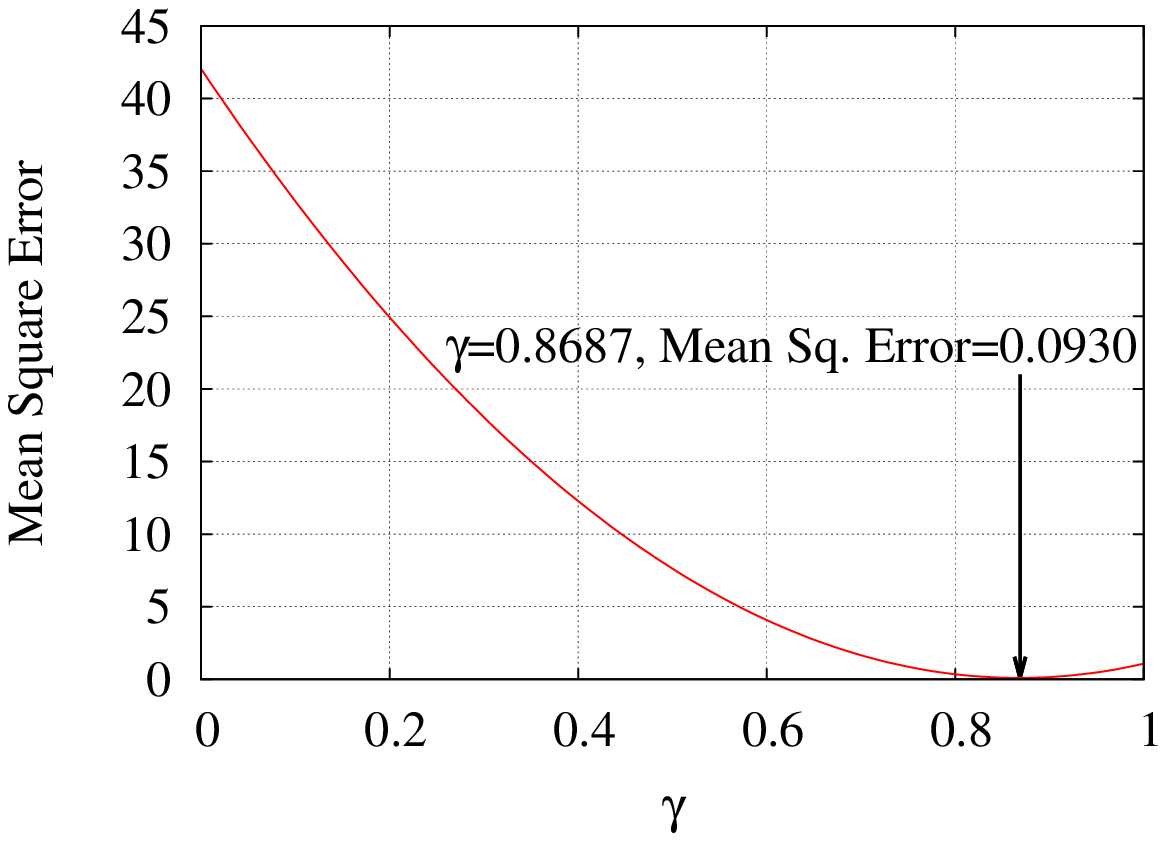}}
  \caption{$\gamma$ that minimizes mean square error}
  \label{fig:gamma}
\end{figure}

\subsection{Determination of $\gamma$}
To determine $\gamma$, we perform a set of experiments to measure MPTCP energy consumption while a device downloads or uploads files of various sizes using MPTCP with WiFi and LTE. We choose $\gamma$ to minimize the mean square error between measured and estimated values.

Figure \ref{fig:gamma} shows the mean square error between measured and estimated energy consumption as a function of $\gamma$ when MPTCP uses WiFi and LTE. As shown in Figure \ref{fig:gamma}, the mean square errors are minimized when $\gamma$ is equal to 0.8485 (for downloads) and 0.8687 (for uploads). Approximately $13\sim16$\% of energy appears to be consumed by shared components when MPTCP is simultaneously operating over both the WiFi and LTE interfaces. One example of shared energy consumption might be CPU processing power to handle MPTCP operations.

In addition, our phone uses the Qualcomm Snapdragon S4 \cite{SamsungSIIhasSnapdragon,SnapdragonS4} with the LTE integrated on the chip, thus, it seems likely that they share some amount of energy while operating.
Since we do not have enough knowledge on the electrical schematic and operation of the mobile device, in the rest of this paper, we use the measured values $\gamma^{d}=0.8485$ (for downloads) and $\gamma^{u}=0.8687$ (for uploads) for our MPTCP energy model. Note that in these experiments, the device uses WiFi and LTE interfaces for downloading and uploading: $\gamma$ can differ in the case when MPTCP uses WiFi and HSDPA interfaces. Also, as with the profiling results in the previous section, using different mobile devices may result in different estimates for $\gamma$. We will explore appropriate values of $\gamma$ for various mobile devices in future work.

\subsection{Model Validation}
\begin{figure}[t!!!]
  \centering
  \subfigure[Downloads]{\includegraphics[scale=0.35]{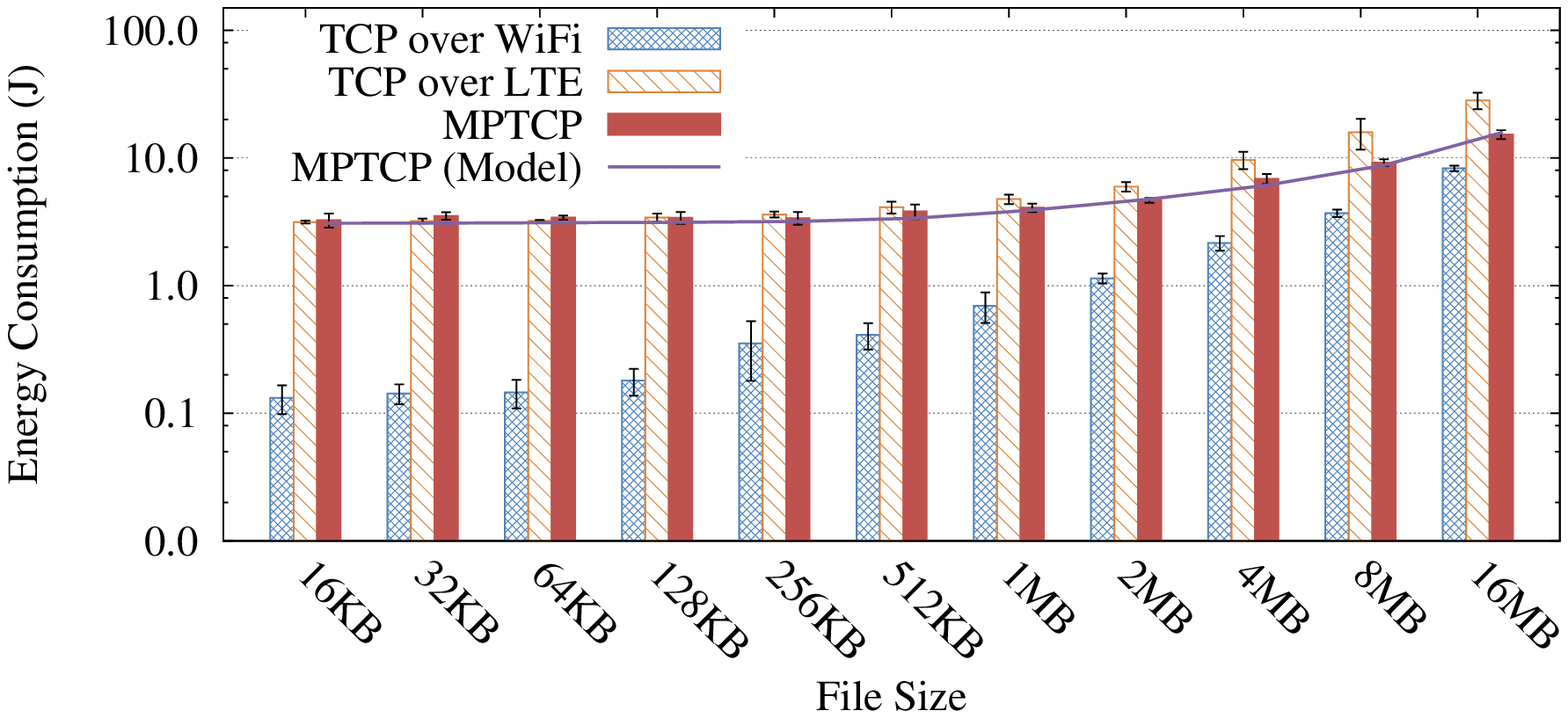}}
  \subfigure[Uploads]{\includegraphics[scale=0.35]{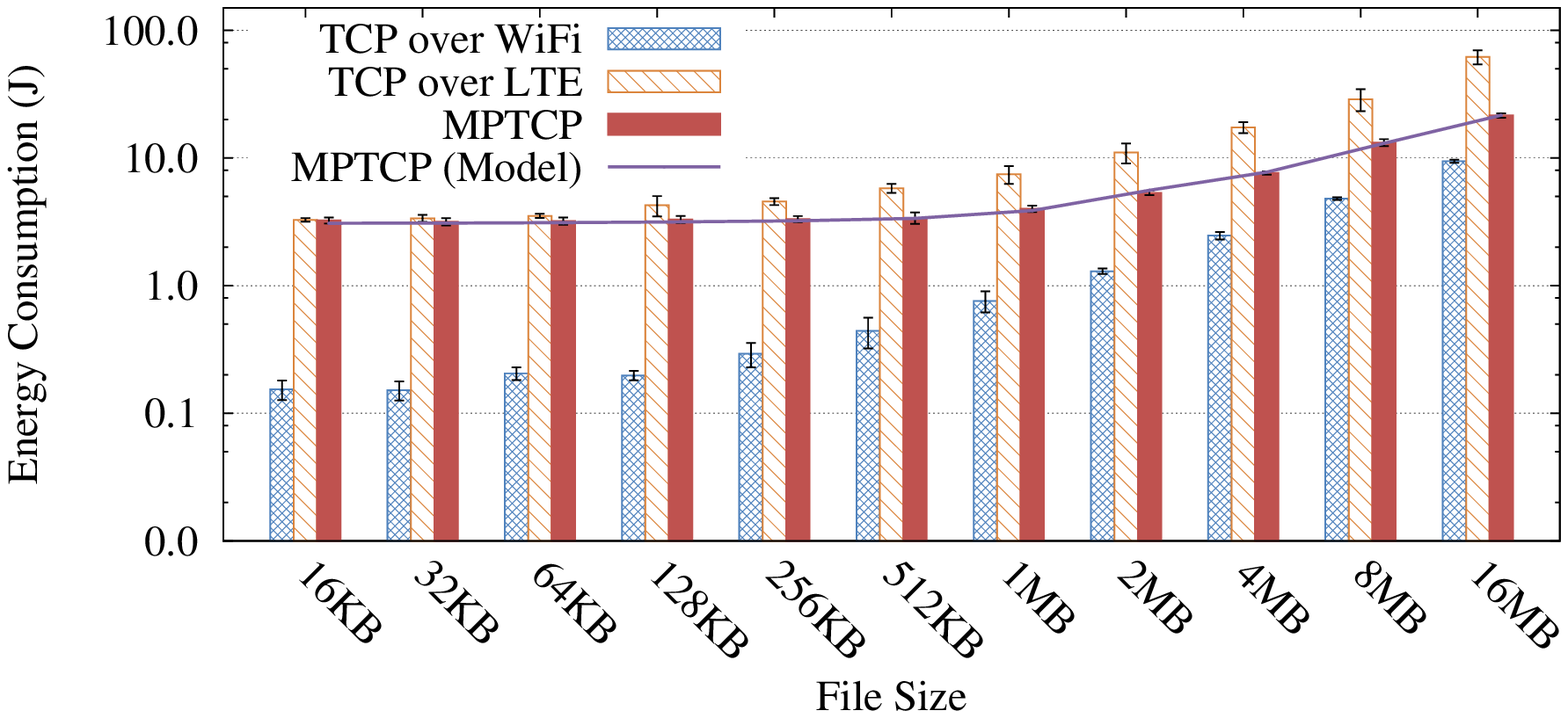}}
  \caption{Total Energy Consumption according to File Size}
  \label{fig:validation}
\end{figure}

\begin{figure}[t!!!]
  \centering
  \includegraphics[scale=0.38]{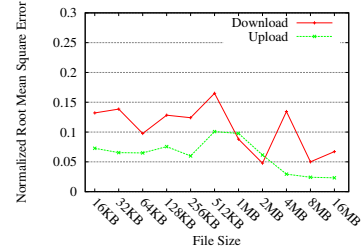}
  \caption{Normalized RMSE}
  \vspace{-0.3cm}
  \label{fig:error}
\end{figure}

\begin{figure*}[t!!!]
  \centering
  \subfigure[1MB Download]{\includegraphics[scale=0.35]{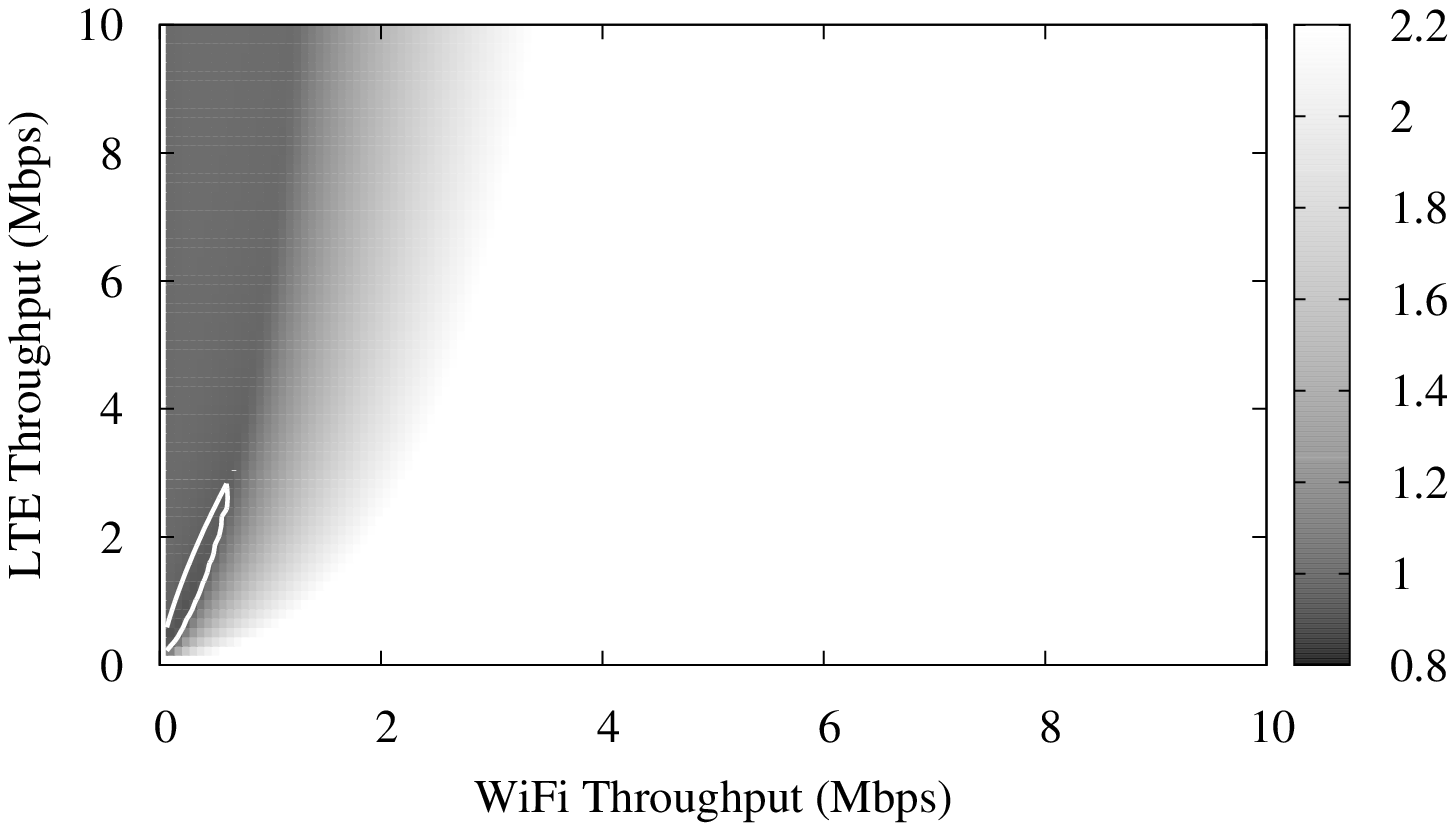}}
  \subfigure[4MB Download]{\includegraphics[scale=0.35]{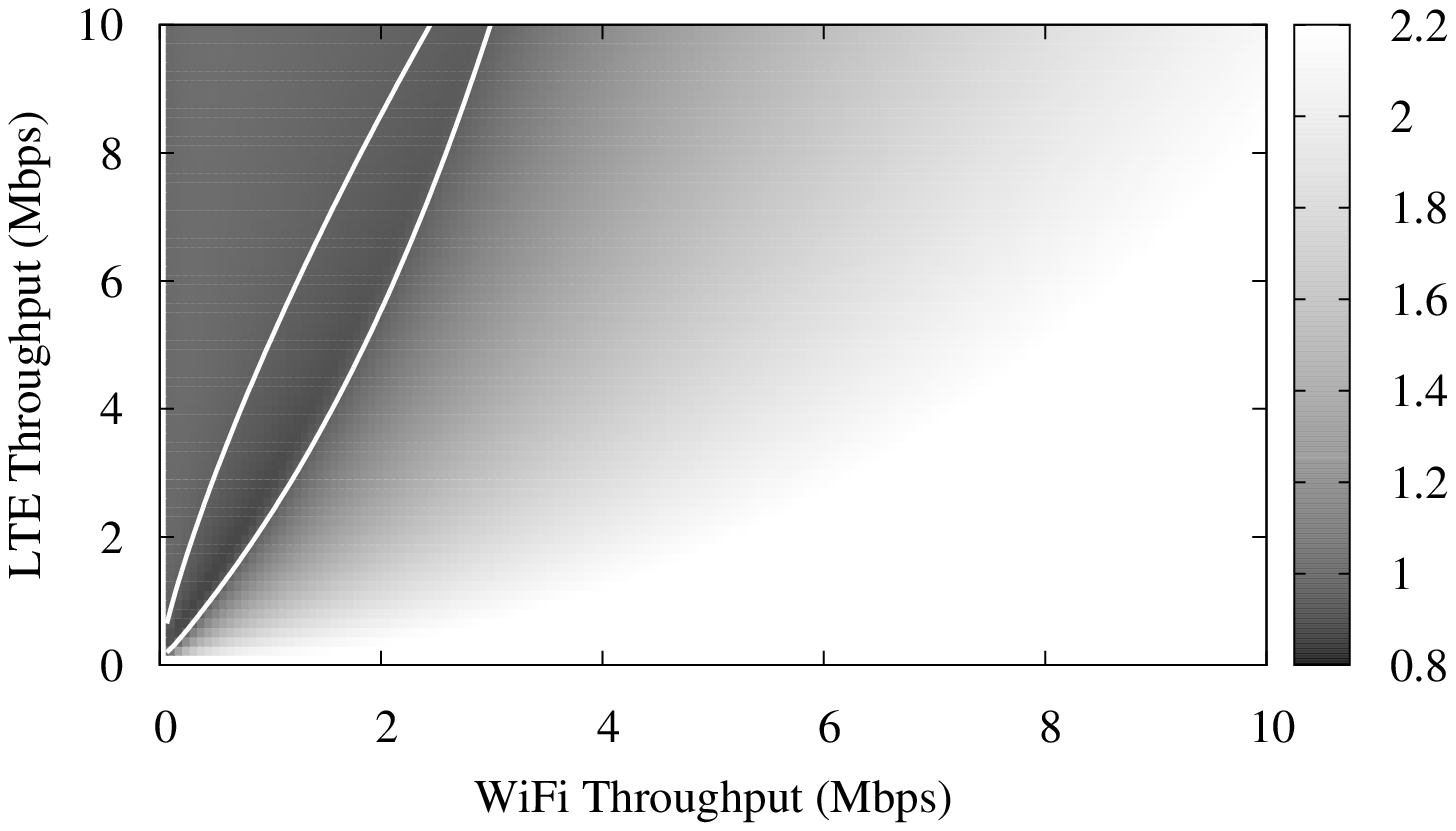}}
  \subfigure[8MB Download]{\includegraphics[scale=0.35]{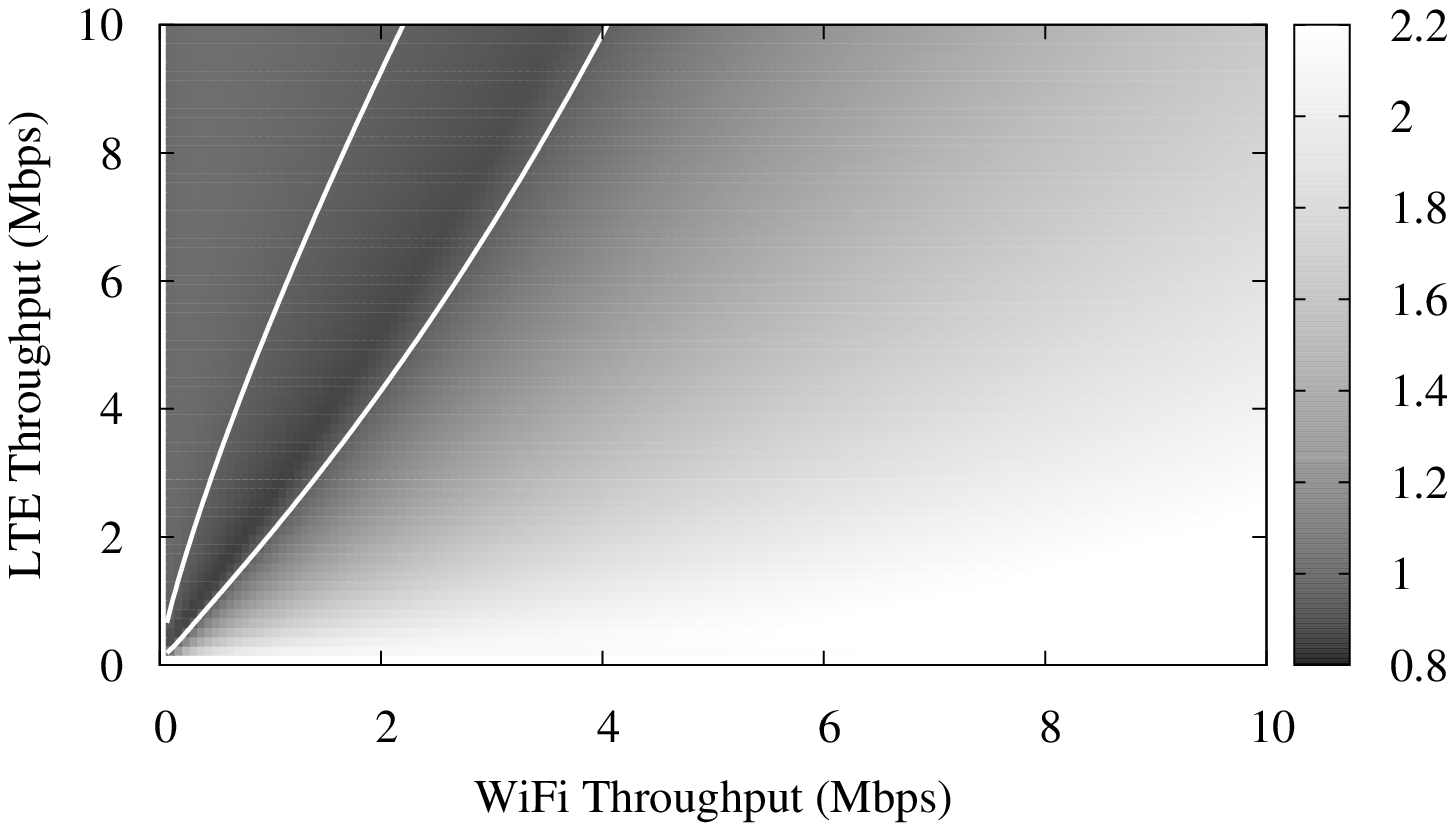}}
  \caption{MPTCP Total Energy Consumption Normalized by Most Energy Efficient TCP}
  \label{fig:norm-mptcp}
\end{figure*}
Figure \ref{fig:validation} shows the average total energy consumption of TCP over each interface (WiFi and LTE) and MPTCP when our device downloads/uploads a file of fixed size ten times each for several different file sizes. We observe that MPTCP is less energy efficient than single-path TCP over WiFi as the file size decreases. This is because of the fixed overhead from the \textit{promotion} and \textit{tail} state of LTE: when a file size is small, the device spends energy to establish an LTE subflow connection even though it rarely transmits packets through LTE. In our experiments, MPTCP does not even utilize an LTE subflow until the transfer size becomes larger than 512KB (effective use of LTE starts even after a transfer becomes larger than 2MB). This is due to the relatively large RTT (about 65 ms for downloads and 95 ms for uploads) over the LTE path; setting up a subflow over LTE takes longer than over WiFi, which has much lower RTT (about 15 ms for downloads and uploads). The larger RTT also means it takes time for the congestion window of the LTE subflow to open up \cite{Chen2013}.
Thus, a small file ($<$512KB) transmission completes in a short time and consumes a relatively small amount of energy for packet transfer compared to the fixed overhead.
Consequently, MPTCP yields similar energy consumption to TCP over LTE when a file is smaller than 512KB, which is larger than that of TCP over WiFi, even though it allocates almost of all traffic to WiFi. We see that in the case of small file transmissions, MPTCP uses power inefficiently, whereas TCP over WiFi is more energy efficient. 

We also calculate the expected MPTCP energy consumption based on our model. As shown in Figure \ref{fig:validation}, our model accurately estimates the energy consumption of MPTCP. 
Figure \ref{fig:error} presents the root mean square errors (RMSE) normalized by the average measured energy consumption for downloading/uploading files of each size. As shown in Figure \ref{fig:error}, the MPTCP energy model estimates differ from the measure values by less than 17\%. As file size increases, the MPTCP energy model becomes more accurate. Note that the large percentage error when the file size is small ($<$1MB) is because the entire energy consumption is relatively small ($<$5J), even though an absolute error is small (error of $<$1J).

\section{Energy Aware MPTCP}
\label{sec:eamptcp}
In this section, we introduce eMPTCP which improves on standard MPTCP by being more energy efficient.
We focus on downloads over the WiFi and LTE interfaces since they are more common.

\subsection{Motivation}
First, based on our MPTCP energy model, we characterize the throughput region where MPTCP is more energy efficient than standard TCP and MPTCP, given the file size for downloads. For example, Figure \ref{fig:norm-mptcp} presents the MPTCP energy consumption estimated for 1MB, 4MB, and 8MB downloads, normalized by the energy consumption of standard TCP over the most energy efficient interface given the achieved throughput over each interface. In Figure \ref{fig:norm-mptcp},
the regions inside white curves correspond to the throughput values where MPTCP is more energy efficient than either TCP over WiFi or TCP over LTE.
At the left side of the region, TCP over LTE is the most energy efficient while TCP over WiFi is the most energy efficient at the right of the region. We have also observed that the operating region where MPTCP is most energy efficient becomes smaller as file size decreases. When file sizes are smaller than 1MB, MPTCP is more efficient than standard TCP only in a significantly limited region and TCP over WiFi is better than MPTCP even when the available WiFi throughput is small. Inspired by this finding, in Section \ref{subsec:delayed}, we propose a delayed LTE subflow establishment mechanism that allows MPTCP to reduce unnecessary energy consumption due to the fixed overhead while downloading a small file.

\begin{figure}[t]
  \centering
  \includegraphics[scale=0.38]{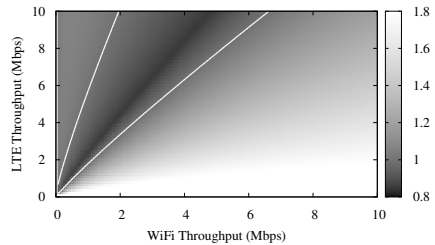}
  \vspace{-0.3cm}
  \caption{Energy Consumption per Downloaded Byte of Using WiFi and LTE Normalized by Most Energy Efficient TCP}
  \vspace{-0.1cm}
  \label{fig:norm-mptcp-detail-transferred-byte}
\end{figure}

We also investigate the energy efficiency of using both WiFi and LTE interfaces according to available throughput during packet transfer. Figure \ref{fig:norm-mptcp-detail-transferred-byte} presents the energy consumption per downloaded byte over both WiFi and LTE, normalized by that achieved using the best single interface. As with Figure \ref{fig:norm-mptcp}, using both interfaces consumes the smallest amount of energy to download a byte inside the white curves. This shows that we should be careful to choose the appropriate combination of interfaces in order to save energy according to available throughput of each interface. Based on this result, we propose a subflow usage management algorithm in Section \ref{subsec:mgmt}.

\subsection{Delayed LTE Subflow Establishment} \label{subsec:delayed}
The delayed subflow management algorithm is shown in Algorithm \ref{alg:1} with notation defined in Table \ref{tab:notation}.
Algorithm \ref{alg:1} is executed only when eMPTCP initiates a new connection.
We assume that a device uses the WiFi interface as the primary network interface.
In this case, the device tries to set up an LTE subflow only some time after establishing an WiFi subflow.
When the file to be downloaded is small, we want to avoid the unnecessary expenditure of energy to establish an LTE subflow connection.
To avoid such an unnecessary LTE subflow establishment, eMPTCP introduces a delay between WiFi and LTE subflow establishment: eMPTCP does not start the LTE subflow until it receives $\kappa$ bytes through the WiFi interface.
Using the delayed LTE subflow establishment, eMPTCP does not consume energy for the \textit{promotion} and \textit{tail} states of LTE interface when a device downloads a file smaller than $\kappa$ bytes.
However, if the available WiFi throughput is extremely small, using only the WiFi subflow until the threshold $\kappa$ can incur more energy consumption than using both (see Figure \ref{fig:norm-mptcp}).
For example, if the WiFi bandwidth is smaller than 3Mbps and the LTE bandwidth is larger than 10Mbps when downloading a 8MB file, using both is better.
Therefore, to prevent using only a slow WiFi subflow, eMPTCP uses a timer to trigger LTE subflow establishment: if the timer expires after $\tau$ seconds, eMPTCP establishes an LTE subflow even though the downloaded amount through WiFi does not reach the threshold $\kappa$.

\begin{table}[t]
  \centering
  \caption{Notations}\label{tab:notation}
  \begin{small}
  \begin{tabular}{c|m{6.5cm}}
  \hline
   Symbol & {\centering Definition} \\
  \hline
  $\kappa$ & Download Amount Threshold for delaying LTE subflow establishment \\
  $\tau$ & Timer Threshold for delaying LTE subflow establishment (ms) \\
  $\delta$ & Bandwidth Estimation Interval (ms) \\
  $h$ & Holt-Winters step ahead parameter \\
  $\rho$ & Holt-Winters EWMA parameter \\
  \hline
  \end{tabular}
  \end{small}
\end{table}

\begin{algorithm}[t]
\begin{small}
\begin{algorithmic}
	\If{MPTCP starts a new connection with WiFi}
		\State Postpone LTE subflow establishment
		\State Trigger a timer that expires in $\tau$ ms
	\Else
		\State Establish all remaining subflows
	\EndIf
	\While{delaying LTE subflow establishment}
		\If{Download more than $\kappa$ bytes or Timer is expired}
			\State Establish LTE subflow
		\EndIf
	\EndWhile
\end{algorithmic}
\caption{Delayed LTE subflow Establishment}
\label{alg:1}
\end{small}
\end{algorithm}

\begin{algorithm}[t]
\begin{small}
\begin{algorithmic}
\Function{Holt-Winters\_BW\_Predictor}{$Y_i$, $h$}
	\If{$i \leq 2$}
		\State $a \leftarrow Y_i$
		\State $b \leftarrow Y_{i-1}$
	\Else
		\State $temp \leftarrow a$
		\State $a \leftarrow \rho \times Y_i + (1-\rho) (a + b)$
		\State $b \leftarrow \rho \times (a-temp) + (1-\rho) b$
	\EndIf
	
	\State Return $a + bh$ /* return h-step ahead prediction */
\EndFunction
\end{algorithmic}
\caption{Holt-Winters Bandwidth Predictor}
\label{alg:2}
\end{small}
\end{algorithm}

\begin{algorithm}[t]
\begin{small}
\begin{algorithmic}
\For{every $\delta$ ms}
    \State $i \leftarrow i+1$
    \State $Y_i = B / \delta$ /* $B$: Downloaded bytes for $\delta$ ms via subflow */
    \State $BW =$ Holt-Winters\_BW\_Predictor($Y_i$, h)
    \If{subflow is associated with WiFi interface}
    		\State $WiFi\_BW = BW$
    	\Else
    		\State $LTE\_BW = BW$
    	\EndIf
    \If{$WiFi\_BW$ and $LTE\_BW$ in WiFi-only region}
        \State Suspend LTE subflow
    \Else
        \State Resume LTE subflow
    \EndIf
\EndFor
\end{algorithmic}
\caption{Subflow Management}
\label{alg:3}
\end{small}
\end{algorithm}

\subsection{Subflow Usage Management} \label{subsec:mgmt}
After establishing an MPTCP connection, eMPTCP uses the subflow management algorithm shown in Algorithm \ref{alg:3}.
The algorithm decides whether to use both interfaces or WiFi-only for data transfer based on the estimated available throughput of each interface.
Note that eMPTCP does not switch to using LTE-only since its expected gain is not much more than using both interfaces, as shown in Figure \ref{fig:norm-mptcp-detail-transferred-byte}.
Thus, the right white curve in Figure \ref{fig:norm-mptcp-detail-transferred-byte} determines the throughput thresholds for eMPTCP to switch between using both and WiFi-only.
To estimate current available throughput, eMPTCP samples downloaded bytes through each interface every $\delta$ ms.
Given the sampled throughputs, to predict bandwidth changes, eMPTCP uses a Holt-Winters time-series forecasting algorithm \cite{Holt-Winters}, which is known as a more accurate predictor than formula-based predictors using a function of underlying path characteristics \cite{sigcomm-holtwinters}.
The Holt-Winters algorithm is shown in Algorithm \ref{alg:2}.

When a device decides to switch from using both interfaces to WiFi-only or vice versa, this decision needs to be communicated to the sender-side. To inform the sender side of the state change, we add an MP\_PRIO option \cite{rfc6824}, which changes the priority of the LTE subflow, to the next packet to be transmitted. 

When the sender needs to switch from using WiFi-only to using both interfaces, eMPTCP at the sender needs utilize the LTE subflow quickly. To this end, eMPTCP disables CWND reset after an idle period longer than the retransmission timeout in RFC2861 \cite{RFC_2861} to ensure that an LTE subflow avoids unnecessary slow-start when eMPTCP starts re-using the LTE subflow. Also, eMPTCP sets the measured round trip time (RTT) of the LTE subflow to zero when it releases the low priority status of LTE subflow. This modification enables an LTE subflow to be quickly probed by the MPTCP subflow scheduler, since it selects a subflow with the lowest RTT for packet transmission \cite{Raiciu2012}.

\section{Evaluation}
\label{sec:evaluation}
In this section, we evaluate eMPTCP in terms of energy consumption and download performance, comparing it to standard MPTCP and TCP over WiFi.
We consider three experimental scenarios where we change the WiFi bandwidth, the background traffic, and the presence of mobility, which are described in more detail below.

\subsection{Setup}
We deploy the same setup used in our single-path TCP profiling measurements in Section \ref{sec:setup}.
We set eMPTCP parameters as follows: the download amount threshold $\kappa$ to postpone an LTE subflow is set to 1MB since MPTCP is rarely more energy efficient than single path TCP when downloading a file smaller than 1MB as shown in Figure \ref{fig:norm-mptcp}(a).
The timer threshold $\tau$ is set to 3 seconds.
We set the polling interval $\delta$ to sample throughput every 200 ms, which yields the best energy performance in our preliminary experiments with different values of $\delta$. eMPTCP uses a one step ahead predicted value with $\rho=0.125$ from the Holt-Winters forecasting algorithm to decide whether to use both interfaces or WiFi-only. 

\begin{figure}[t!!!]
  \centering
  \subfigure[Energy Consumption]{\includegraphics[scale=0.32]{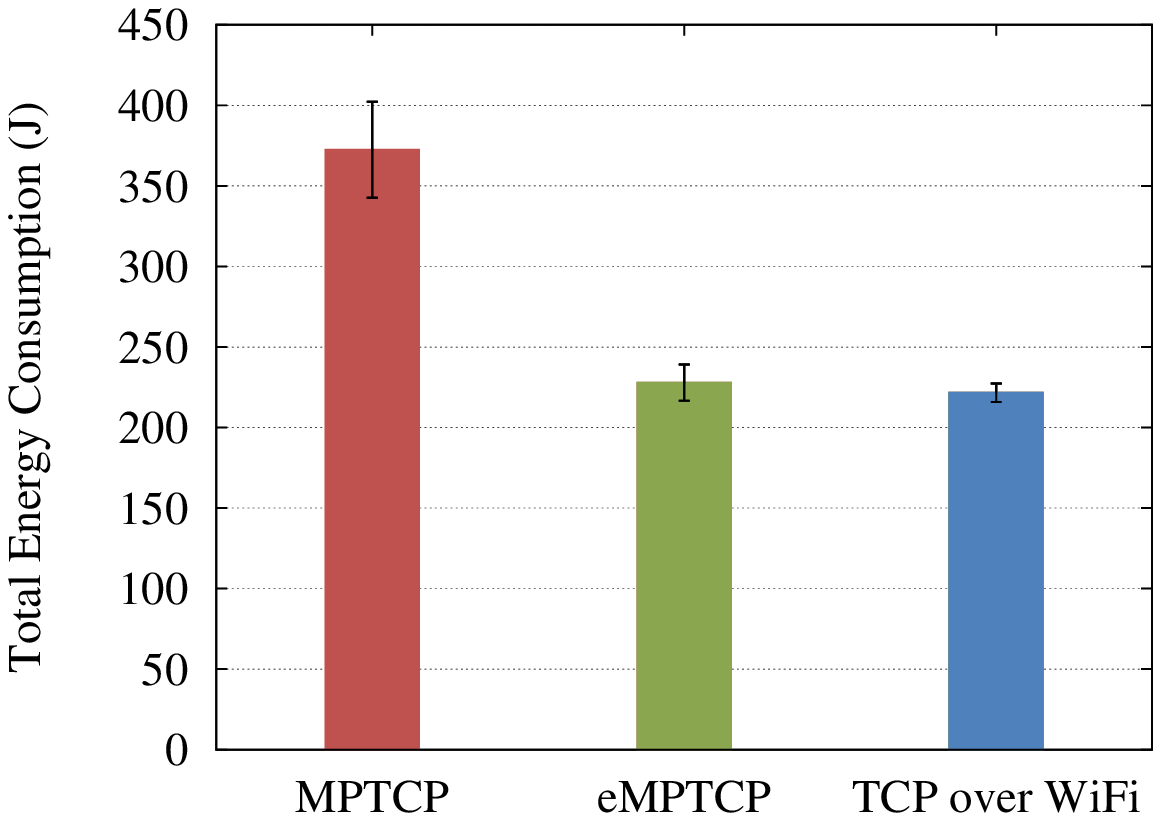}}
  \subfigure[Download Time]{\includegraphics[scale=0.32]{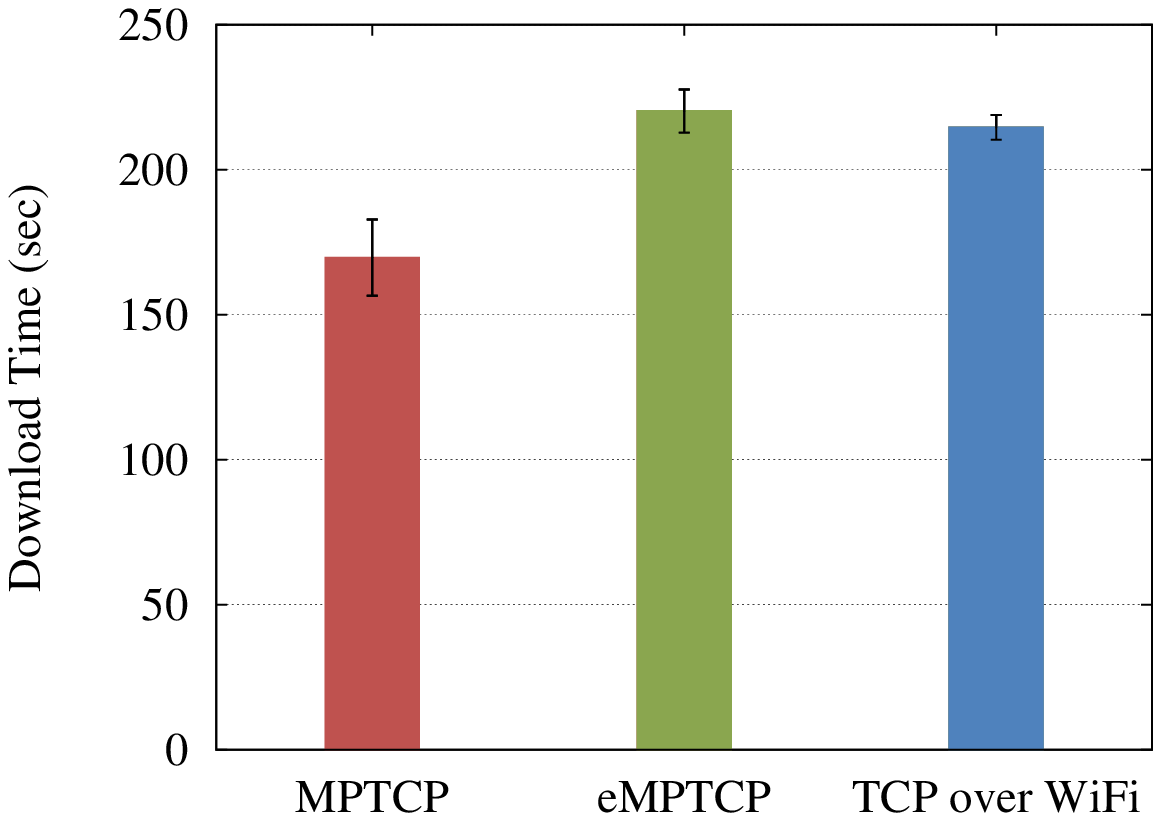}}
  \caption{Persistent High WiFi Bandwidth}
  \label{fig:mgmt-comp-stable-fast}
\end{figure}

\begin{figure}[t!!!]
  \centering
  \subfigure[Energy Consumption]{\includegraphics[scale=0.32]{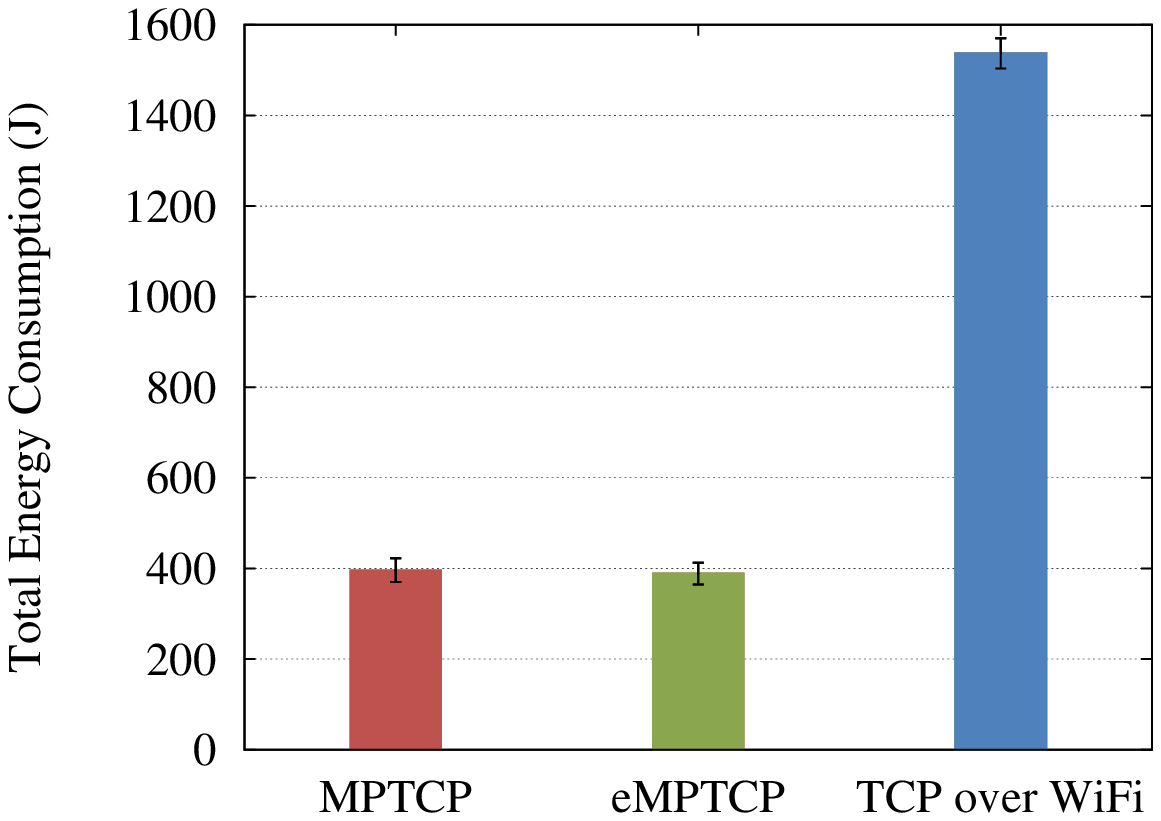}}
  \subfigure[Download Time]{\includegraphics[scale=0.32]{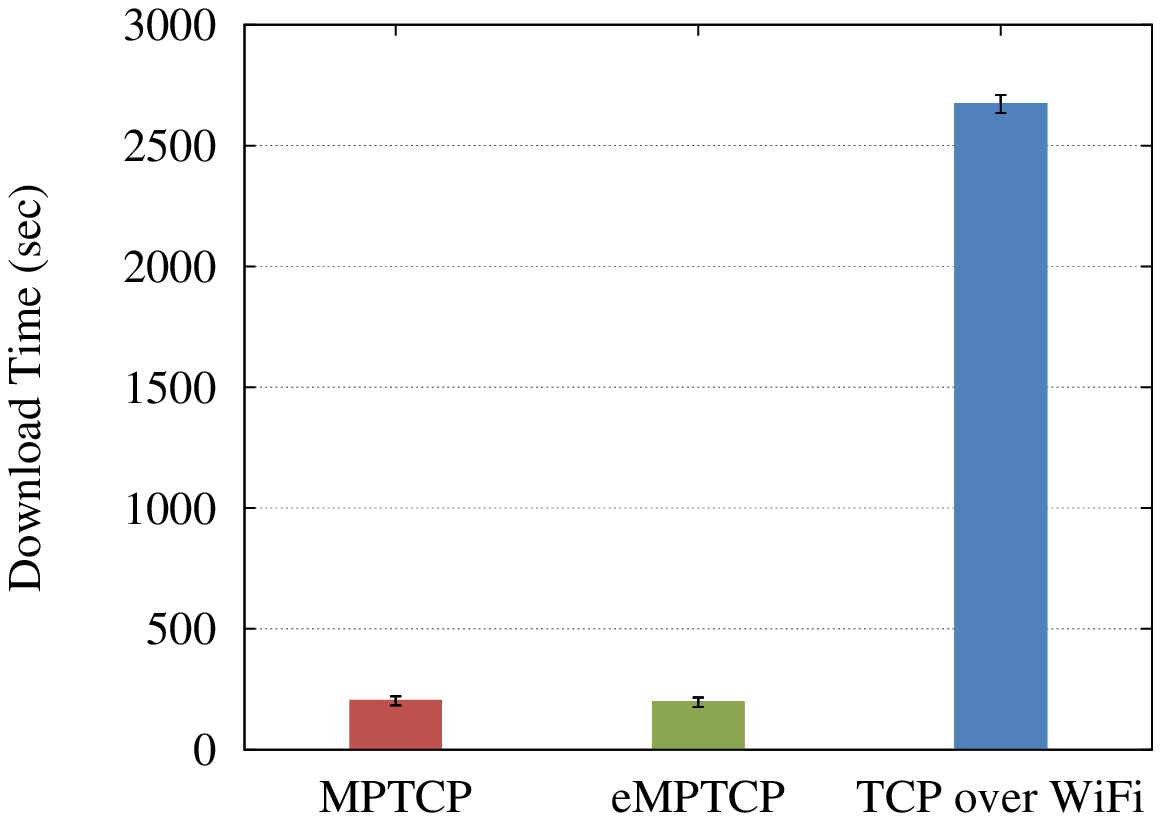}}
  \caption{Persistent Low WiFi Bandwidth}
  \label{fig:mgmt-comp-stable-slow}
\end{figure}

\subsection{Experiments with Static Configuration}

The purpose of these experiments is to show that, in relatively simple static environments, eMPTCP makes the proper path decisions to optimize power consumption.
We measure energy consumption and download time of eMPTCP, MPTCP, and TCP over WiFi for two extreme cases: persistent high ($>$10Mbps) and low ($<$1Mbps) WiFi bandwidth while the device downloads a 256MB file at a static location.
Figures \ref{fig:mgmt-comp-stable-fast} and \ref{fig:mgmt-comp-stable-slow} compare the energy consumption and download times averaged over five runs for the two cases.
In the first case, WiFi bandwidth is high, and thus using it is more power efficient than using LTE or both interfaces.
Figure \ref{fig:mgmt-comp-stable-fast} shows that eMPTCP chooses WiFi-only, effectively behaving like TCP over WiFi.
In contrast, when the WiFi bandwidth is small ($<$1Mbps), and thus less energy-efficient as LTE, Figure \ref{fig:mgmt-comp-stable-slow} shows that eMPTCP yields almost the same performance as MPTCP by using both interfaces (after the LTE startup delay determined by parameters $\kappa$ and $\tau$).
This illustrates that eMPTCP seeks the most energy efficient path usage, while preserving MPTCP's robustness to path degradation, and without user involvement.

\subsection{Experiments with Bandwidth Changes}

\begin{figure}[t!!!]
  \centering
  \includegraphics[scale=0.35]{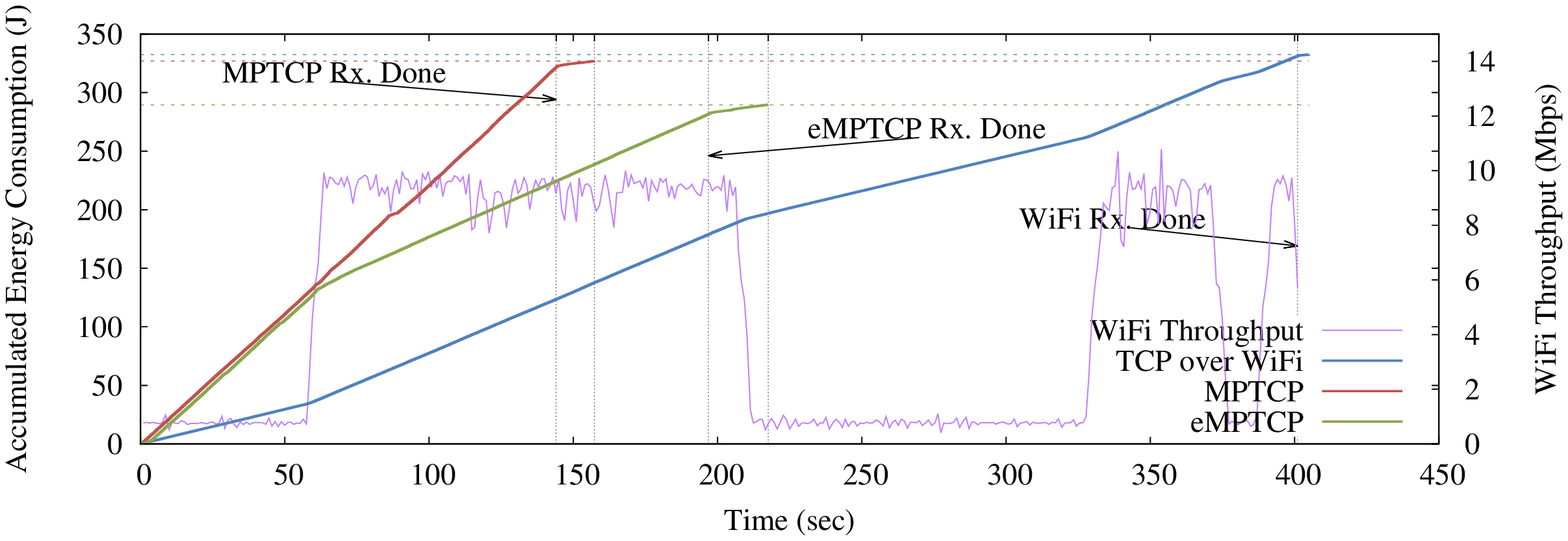}
  \caption{Accumulated Energy Consumption Example with Random WiFi Bandwidth Change}
  \label{fig:mgmt-example-256m-random}
\end{figure}

\begin{figure}[t!!!]
  \centering
  \subfigure[Energy Consumption]{\includegraphics[scale=0.32]{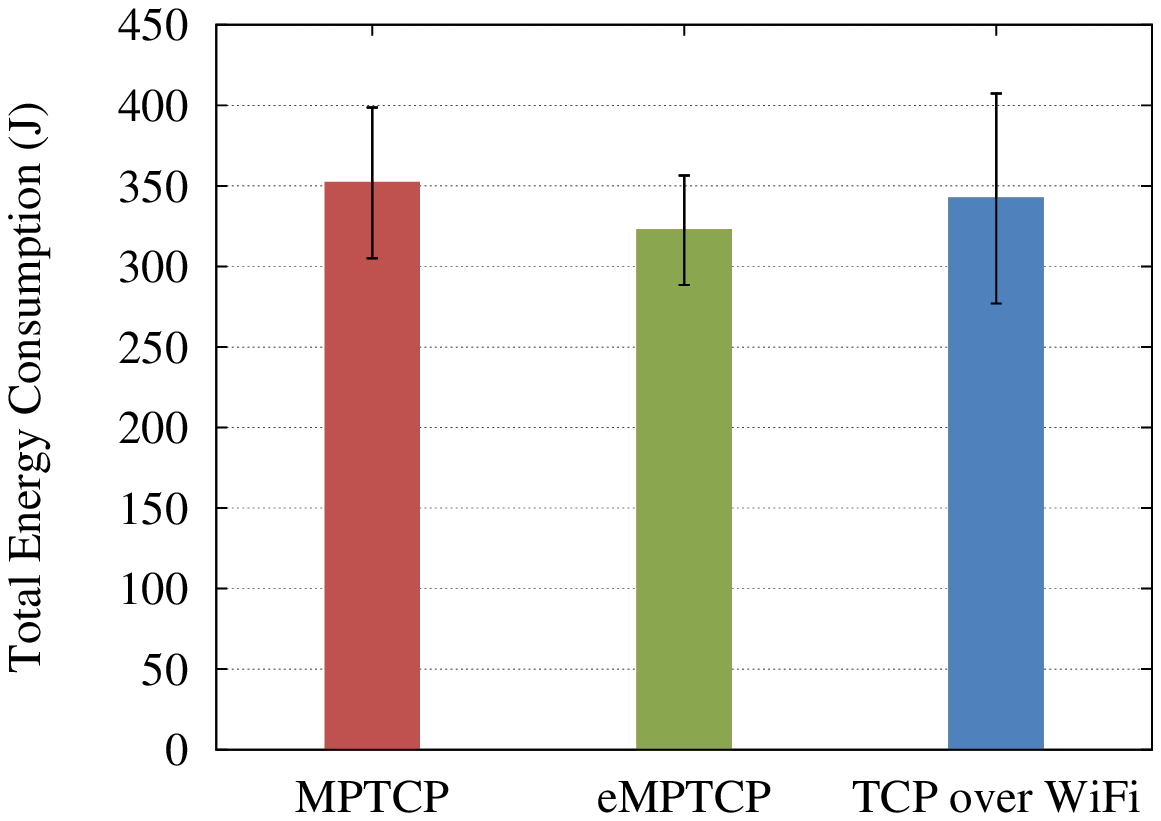}}
  \subfigure[Download Time]{\includegraphics[scale=0.32]{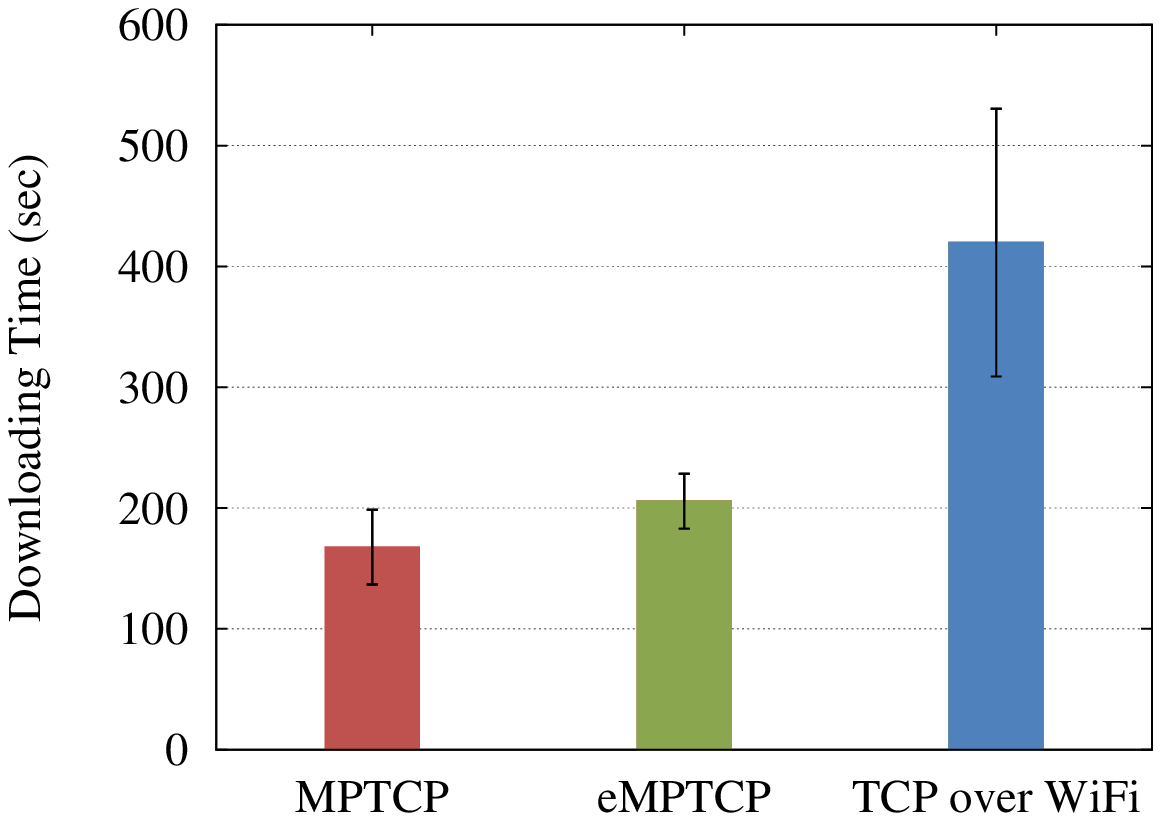}}
  \caption{Random WiFi Bandwidth Changes}
  \label{fig:mgmt-comp-random}
\end{figure}

The goal of these experiments is to examine how robust eMPTCP is to changes in network bandwidth, which can exhibit variability.
In these experiments, link bandwidth changes while the device downloads a 256MB file at a static location.
To simulate available throughput changes, since we do not have any control of the LTE network, we only change the available bandwidth of WiFi by randomly setting the 802.11 physical layer bit rate of our AP between 1Mbps and 18Mbps, where the achieved TCP bandwidth are $<$1Mbps and $>$10Mbps, respectively.

Here, we explore scenarios with random WiFi bandwidth changes where the time between WiFi bandwidth changes is exponentially distributed with a mean of 40 seconds.
Figure \ref{fig:mgmt-example-256m-random} presents an example trace of accumulated energy consumption of eMPTCP, MPTCP, and TCP over WiFi with random WiFi bandwidth changes. At the beginning, eMPTCP uses both interfaces after the LTE startup delay since WiFi throughput is too low to be energy efficient. However, eMPTCP suspends the LTE subflow after WiFi bandwidth increases while MPTCP continues to use both interfaces. By suspending the LTE subflow, through which more energy cost to complete the download is expected even with additional bandwidth, eMPTCP spends less energy but completes the download later than MPTCP. Compared with TCP over WiFi, eMPTCP completes the download sooner and consumes less energy.

Figure \ref{fig:mgmt-comp-random} compares energy consumption and download time averaged over ten runs. In this scenario, eMPTCP consumes approximately 8\% and 6\% less energy than MPTCP and TCP over WiFi, respectively. However, eMPTCP is approximately 22\% slower than MPTCP since it utilizes an LTE subflow only when an LTE subflow can provide an energy gain with additional throughput. In contrast, by utilizing an LTE subflow when energy gain is available (WiFi thoughput is $<$1Mbps), eMPTCP completes downloads twice as fast as TCP over WiFi and consumes less energy. This shows that eMPTCP offers the robustness of MPTCP while obtaining greater energy efficiency than MPTCP. Note that the performance gain achieved by eMPTCP differs according to how WiFi bandwidth changes. If WiFi bandwidth is frequently changing, the switching overhead in eMPTCP may become noticeable: an LTE interface triggers another \textit{promotion} and \textit{tail} states when it starts to be used again.

\subsection{Experiments with Background Traffic} \label{subsec:exp-bgt}
It is well known that multiple WiFi nodes can contend for the air channel, causing interference and loss (e.g., \cite{Manweiler2011}).
The goal of this next section is to see how well eMPTCP copes with with random background traffic.
Background traffic causes available throughput changes similar to link bandwidth changes, but also results in contention and interference in the communication channel.
In these experiments, we utilize two or three interfering nodes, which use the same WiFi channel as the mobile device.
While the device downloads a 256MB file at a static location, each node turns UDP traffic on and off for a random duration,
which is exponentially distributed with a rate of $\lambda_{\text{on}}$ and $\lambda_{\text{off}}$.
We fix $\lambda_{\text{on}}=0.05$, and then perform experiments with $\lambda_{\text{off}}=0.025$ and $\lambda_{\text{off}}=0.05$.
As with the previous scenario, we control background traffic for the WiFi channel only.

\begin{figure}[t!!!]
  \centering
  \includegraphics[scale=0.35]{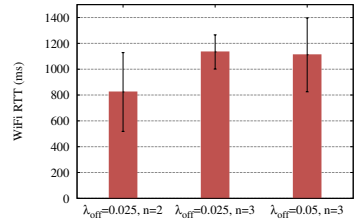}
  \caption{Round Trip Time of TCP over WiFi according to Background Traffic Setup}
  \label{fig:mgmt-inteference-rtt}
\end{figure}

Figure \ref{fig:mgmt-inteference-rtt} presents the average round trip time (RTT) of TCP over WiFi, which is one indicator of the interference level in the WiFi channel, with different numbers of interfering nodes ($n$) and different $\lambda_{\text{off}}$. While larger $n$ apparently increases the average WiFi RTT, $\lambda_{\text{off}}$ does not significantly affect the average RTT (when comparing the cases of $\lambda_{\text{off}}=0.025$ and $\lambda_{\text{off}}=0.05$ with $n=3$). This is because 20 more seconds per individual period without background traffic ($1/0.025 - 1/0.05$) might not be long enough to change the average RTT. However, we can expect the device to obtain more WiFi throughput as background traffic periods become longer (with smaller $\lambda_{\text{off}}$).

\begin{figure}[t!!!]
  \centering
  \subfigure[MPTCP]{\includegraphics[scale=0.34]{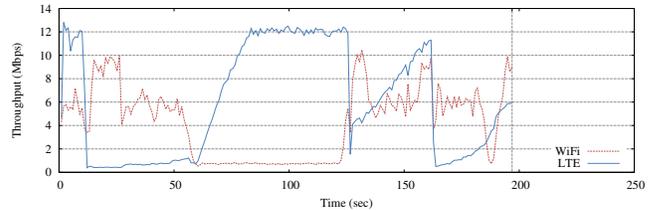}}
  \subfigure[eMPTCP]{\includegraphics[scale=0.34]{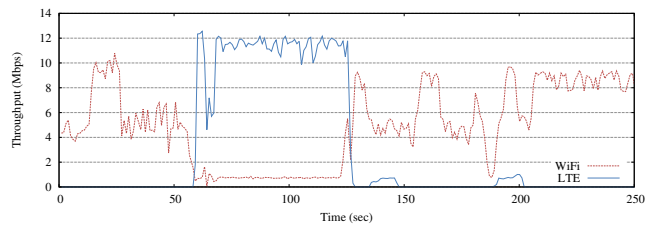}}
  \caption{Example Throughput Trace with Random WiFi Background Traffic ($n=2$, $\lambda_{\text{on}}=0.05$, and $\lambda_{\text{off}}=0.025$)}
  \label{fig:mgmt-comp-example}
\end{figure}

Figure \ref{fig:mgmt-comp-example} shows example throughput traces of MPTCP and eMPTCP when two interfering nodes turn on and off traffic with $\lambda_{\text{on}}=0.05$ and $\lambda_{\text{off}}=0.025$. We observe that MPTCP itself is likely to avoid too aggressive use of an LTE subflow while an WiFi subflow can provides high bandwidth, e.g., at around time 10$\sim$60 seconds in Figure \ref{fig:mgmt-comp-example}(a). This is because the MPTCP subflow scheduler chooses a subflow for packet transmission that has sufficient CWND and the smallest RTT \cite{Raiciu2012}. However, MPTCP consumes energy utilizing an LTE subflow even with such a small throughput gain. In contrast, eMPTCP suspends an LTE subflow while WiFi bandwidth is sufficiently large in order to avoid energy inefficient path usage. In Figure \ref{fig:mgmt-comp-example}(b), we note that eMPTCP incorrectly determines to use an LTE subflow around 140$\sim$150 and 190$\sim$200 seconds. This is caused by the sudden steep WiFi throughput decreases at 140 and 190 seconds, which result in incorrect throughput predictions of eMPTCP. However, eMPTCP stops using the LTE subflow after obtaining improved throughput estimates.

\begin{figure}[t!!!]
  \centering
  \subfigure[Energy Consumption]{\includegraphics[scale=0.29]{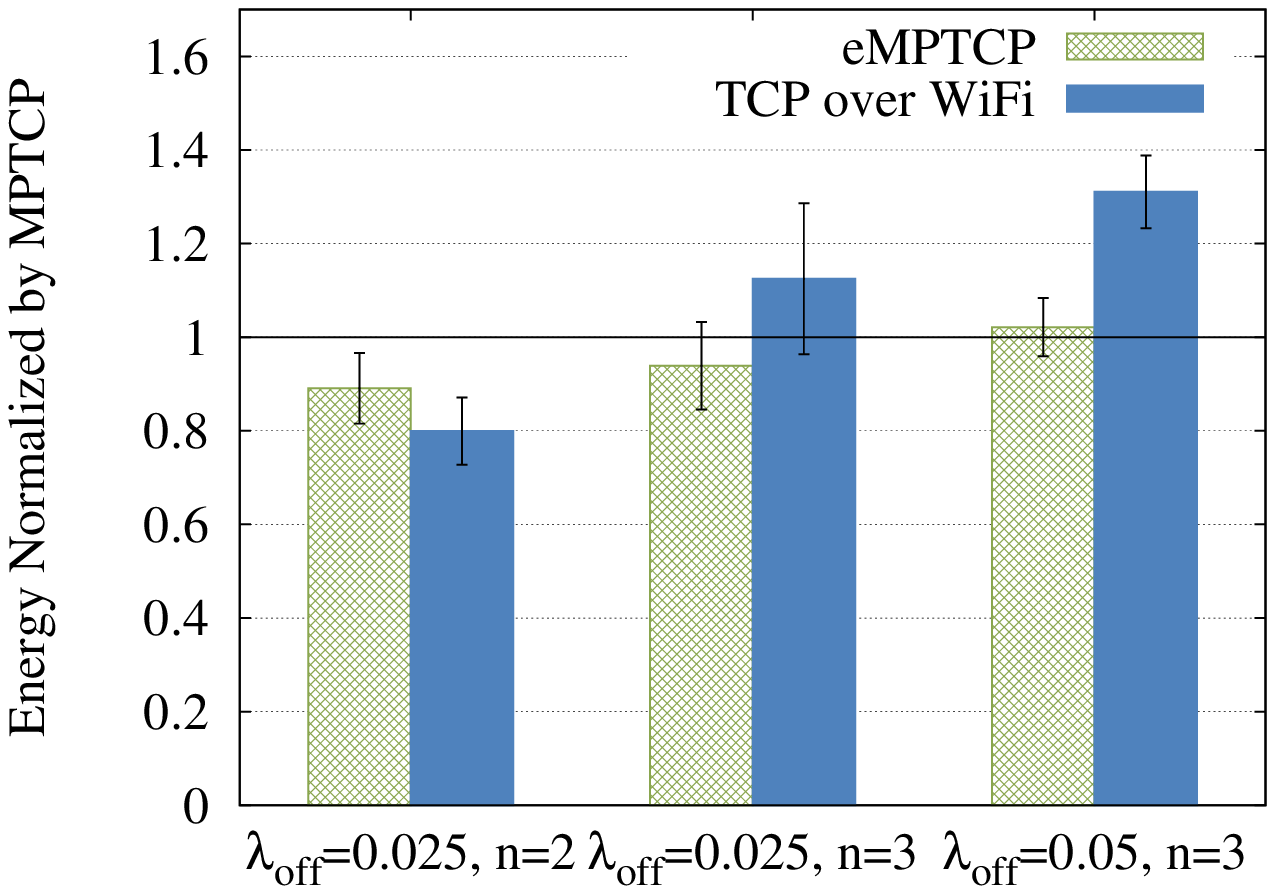}}
  \subfigure[Download Time]{\includegraphics[scale=0.29]{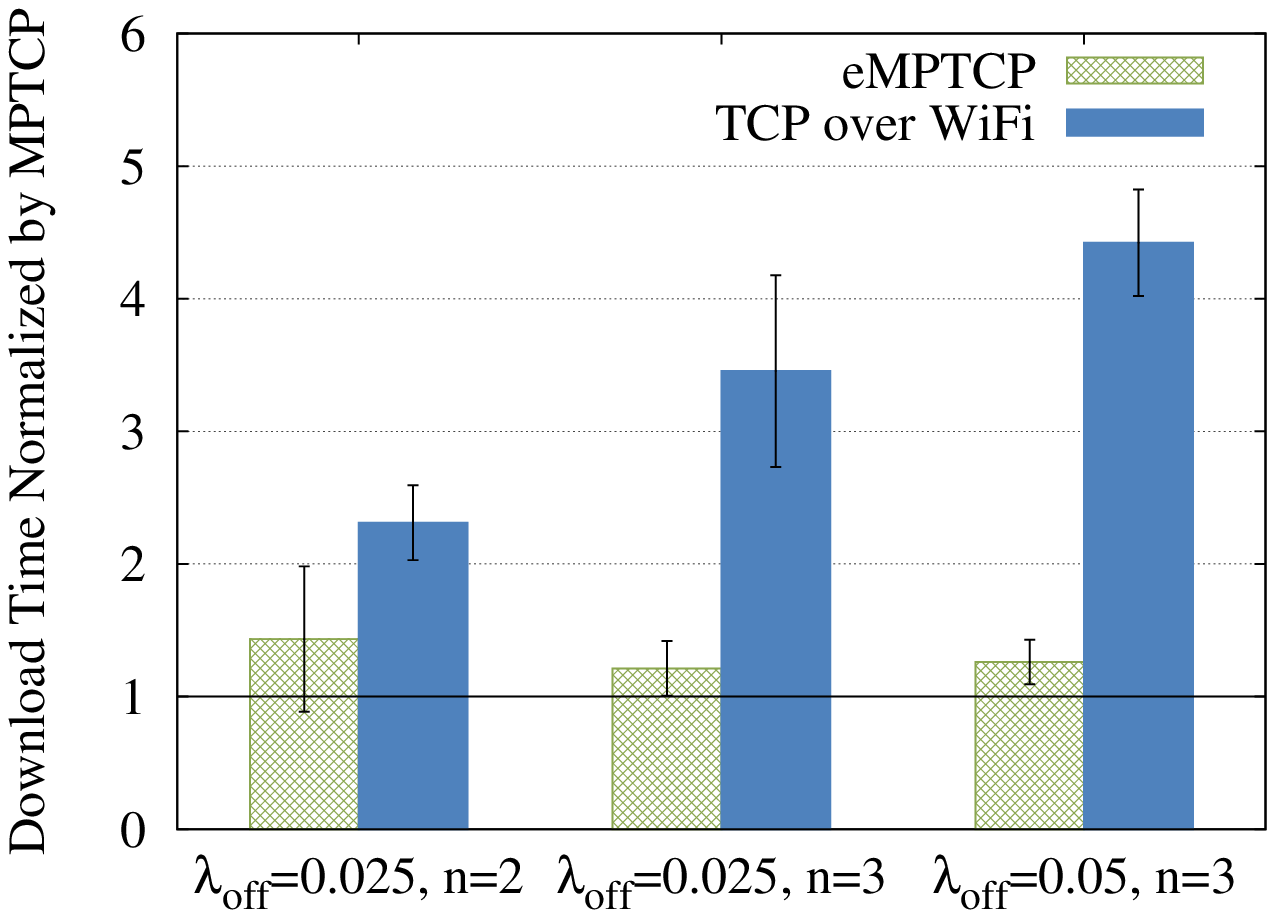}}
  \caption{Random WiFi Background Traffic}
  \label{fig:mgmt-comp-interference-ratio}
\end{figure}

Figure \ref{fig:mgmt-comp-interference-ratio} presents the average energy consumption and download times for different values of $n$ and $\lambda_{\text{off}}$, normalized relative to MPTCP, i.e., smaller numbers are better for both (a) and (b), and when lower than one are better than MPTCP. Values are averaged over five experiments.

As shown in Figures \ref{fig:mgmt-comp-interference-ratio}(a), eMPTCP consumes less energy than MPTCP as $n$ and $\lambda_{\text{off}}$ become smaller. Comparing the cases of $n=2$ and $n=3$ with $\lambda_{\text{off}}=0.025$, we observe that the energy efficiencies of eMPTCP and TCP over WiFi are larger with $n=2$.
Recall that the energy efficiency of eMPTCP is improved when it can suspend an LTE subflow in situations where using WiFi only is more efficient.
Larger numbers of interfering WiFi nodes result in more losses caused by collisions when background traffic is present, resulting in more CWND decreases.
Thus, the device is likely to obtain more TCP bandwidth with a larger CWND when background traffic disappears, resulting in better energy efficiency of TCP over WiFi and eMPTCP.
Note that TCP over WiFi is most energy efficient when $n$ and $\lambda_{\text{off}}$ are equal to 2 and 0.025, respectively.
However, as shown in Figure \ref{fig:mgmt-comp-interference-ratio}(b), in that setting, TCP over WiFi requires twice as much time to complete downloads as eMPTCP, while it consumes just 11\% less energy. We see that eMPTCP is more efficient than TCP over WiFi in terms of the tradeoff between energy consumption and performance.

As shown in Figure \ref{fig:mgmt-comp-interference-ratio}(b), MPTCP provides the best download times, regardless of the values of $n$ and $\lambda_{\text{off}}$, since it always utilizes an LTE subflow. Also, as we expect, the download time of TCP over WiFi becomes significantly larger as $n$ and $\lambda_{\text{off}}$ increase. Compared with MPTCP, eMPTCP requires 20$\sim$40\% more time while consuming 9$\sim$11\% less energy: eMPTCP cannot achieve energy gain as much as the amount of performance degradation. This might be because eMPTCP may sometimes poorly predict bandwidth due to fluctuating throughputs, as shown in Figure \ref{fig:mgmt-comp-example}(b), and it also has additional energy overhead to suspend and resume an LTE subflow due to the \textit{promotion} and \textit{tail} state. In our experiments, the average lengths of period during WiFi background traffic turns off (an LTE subflow is supposed to be suspended) are 20 and 40 seconds. These are not long enough to allow one to ignore energy overhead for \textit{tail} state of LTE that lasts for around 12 seconds.
Compared to TCP over WiFi, eMPTCP obtains a significant performance improvement (up to 70\% less) in download time, while at the same time using less energy.

\subsection{Experiments with Mobility} \label{subsec:exp-mbl}
\begin{figure}[t!!!!]
  \begin{center}
   \includegraphics[scale=0.33]{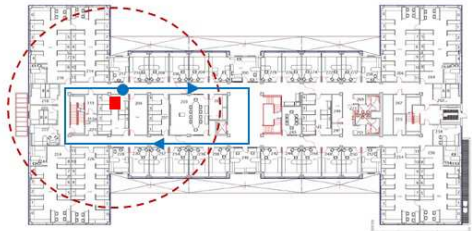}
  \end{center}
  \caption{Mobile Scenario inside the UMass CS Building. Route starts at the blue point. The red square is the AP. The red dashed circle is the estimated usable access range of the AP.}
  \label{fig:map}
\end{figure}

Finally, mobile devices do not simply use wireless networks, they can also move around their environments, causing connectivity changes.
The goal of the experiments in this section is to determine how well eMPTCP performs and adapts in an actual mobile scenario.
We measure energy consumption and download amount while moving for 250 seconds along the route shown in Figure~\ref{fig:map}.
To make our comparison between MPTCP and eMPTCP as fair as possible, we use as similar routes as possible for the experiments.

\begin{figure}[t!!!]
  \centering
  \subfigure[MPTCP]{\includegraphics[scale=0.34]{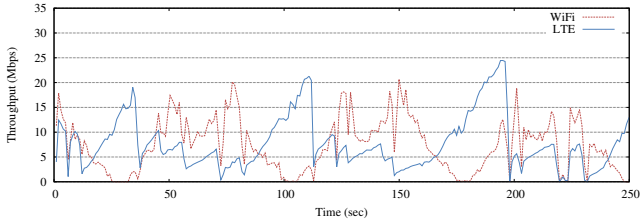}}
  \subfigure[eMPTCP]{\includegraphics[scale=0.34]{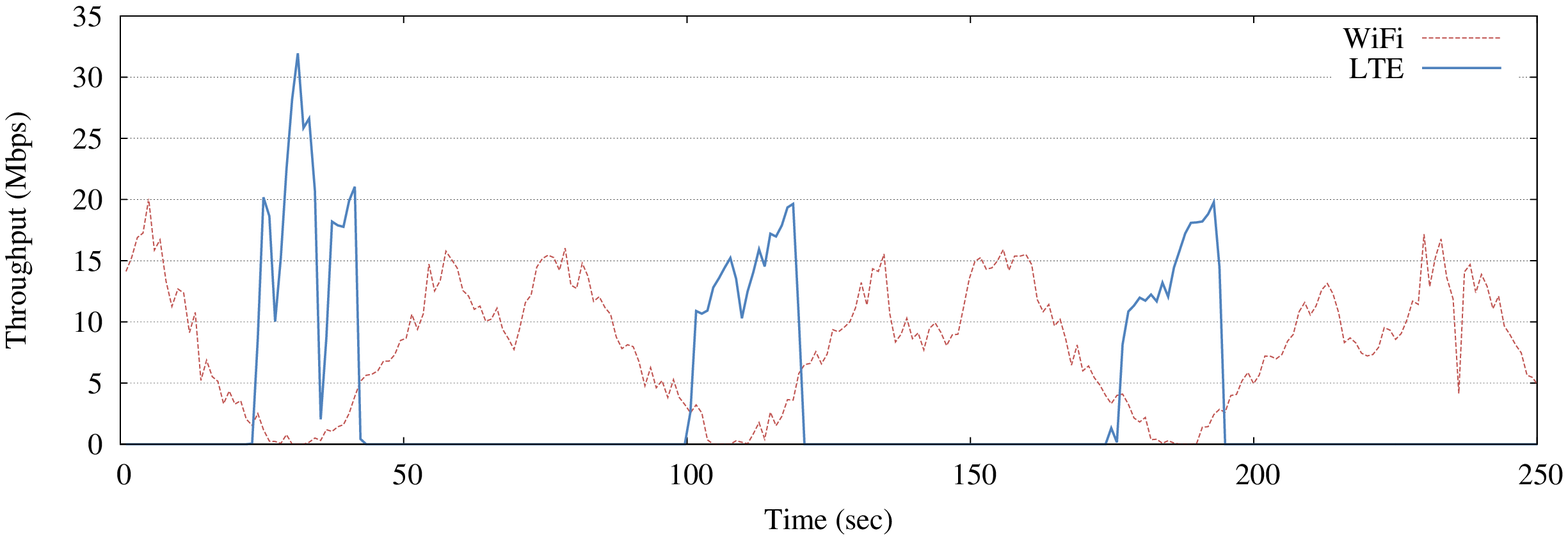}}
  \caption{Example Throughput Trace with Mobile Scenario}
  \label{fig:mgmt-comp-example-mobile}
\end{figure}

\begin{figure}[t!!!]
  \centering
  \includegraphics[scale=0.35]{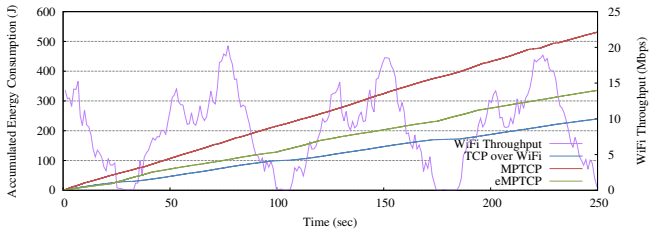}
  \caption{Example of Accumulated Energy Consumption with Mobile Scenario}
  \label{fig:mgmt-example-1024m-mobile}
\end{figure}

Figures \ref{fig:mgmt-comp-example-mobile} and \ref{fig:mgmt-example-1024m-mobile} present example traces of throughput and accumulated energy consumption from our mobile scenario. In this experiment, the device is sometimes within the WiFi communication range, and sometimes outside it, depending on its location. As the device moves outside the WiFi communication range, WiFi throughput decreases, e.g., the duration around 25$\sim$40 seconds in Figure \ref{fig:mgmt-comp-example-mobile}(a) and (b).
At the beginning of the route, MPTCP starts by establishing both subflows, whereas eMPTCP postpones establishing an LTE subflow, since WiFi throughput is high enough to be more energy efficient than using both interfaces.
However, eMPTCP establishes an LTE subflow after the WiFi bandwidth decreases when the device is leaving the WiFi communication range (at around 25 seconds in Figure \ref{fig:mgmt-comp-example-mobile}(b)).
Then, whenever the device cannot obtain enough WiFi bandwidth to be more energy efficient than using both interfaces, eMPTCP utilizes the LTE subflow rather than only using the bad WiFi subflow.
In this experiment, we observe that since the device is inside WiFi communication range most of time, eMPTCP utilizes the LTE subflow only for a few short periods.
Therefore, as shown in Figure \ref{fig:mgmt-example-1024m-mobile}, the slope of eMPTCP's accumulated energy consumption (the energy consumption per second) is larger than that of the slope of TCP over WiFi, but smaller than that of MPTCP.

\begin{figure}[t!!!]
  \centering
  \subfigure[Energy Consumption per Byte]{\includegraphics[scale=0.32]{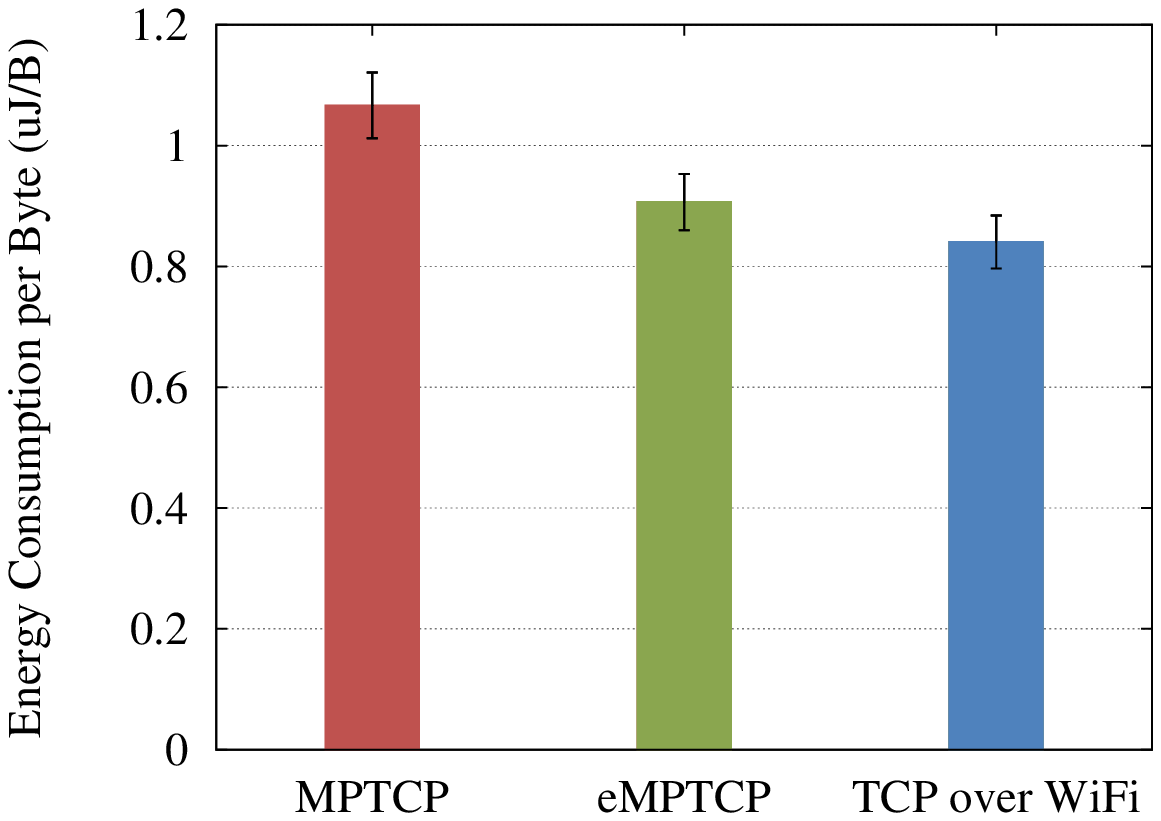}}
  \subfigure[Download Amount]{\includegraphics[scale=0.32]{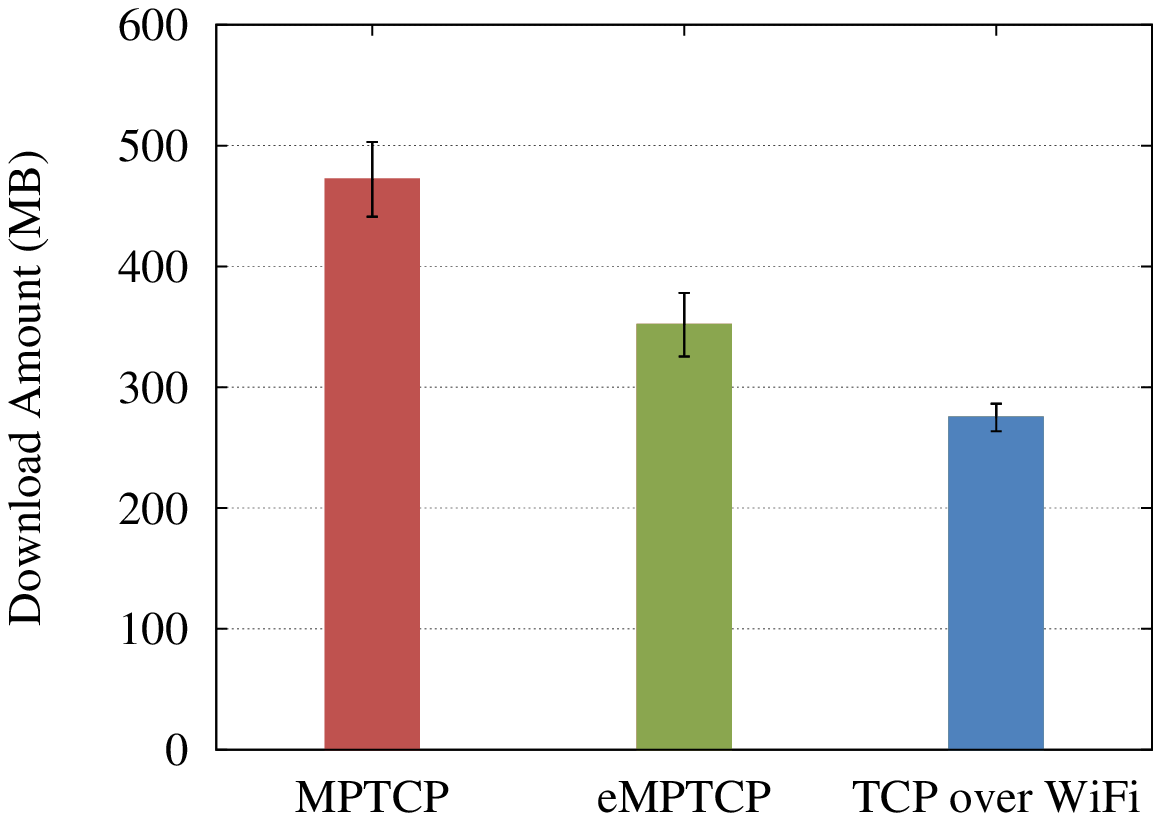}}
  \caption{Mobile Scenario}
  \label{fig:mgmt-comp-mobile}
\end{figure}

We now examine the per-byte energy efficiencies of MPTCP, eMPTCP, and TCP over WiFi.
Figure \ref{fig:mgmt-comp-mobile} compares the energy consumption per byte and download amount for 250 seconds averaged over five runs.
Since eMPTCP utilizes the LTE subflow for only several short periods, its energy consumption per byte is 8\% larger than that of TCP over WiFi and 15\% smaller than that of MPTCP.
Because WiFi throughput degradation is due only to the distance between the AP and device (since there is no WiFi background traffic in these experiments), TCP over WiFi is slightly better in terms of energy efficiency than eMPTCP, similar to the case of $n=2$ and $\lambda_{\text{off}}=0.025$ in Figure \ref{fig:mgmt-comp-interference-ratio}(a).

Comparing the download amounts during the experiments, we observe that eMPTCP downloads 28\% more data than TCP over WiFi even though it yields a similar per-byte energy efficiency to TCP (just 8\% more energy consumption per byte).
As was the case in the experiments with random WiFi background traffic, this result shows that eMPTCP can obtain significant performance gains in the tradeoff between energy consumption and performance, compared to TCP over WiFi.
Recall that eMPTCP consumes 15\% less energy per byte than MPTCP.
With such better per-byte energy efficiency, eMPTCP downloads 25\% less data.
Similar to the results of Figure \ref{fig:mgmt-comp-interference-ratio},
eMPTCP still loses proportionally more in performance than it saves in energy efficiency,
due to the overhead of switching between WiFi-only and using both interfaces.
Improving eMPTCP to achieve better efficiency by considering switching overhead is future work.

\subsection{eMPTCP vs. MPTCP with WiFi First}
Raiciu et.~al \cite{Raiciu2011:Opportunistic} propose a simple strategy called MPTCP with WiFi First, where MPTCP only uses WiFi when available, and only uses the cellular network otherwise.
This is accomplished by placing the cellular subflow in backup mode and activating it only when WiFi is not available.
This simple strategy may seem similar to eMPTCP, however, it cannot take advantage of more dynamic situations where TCP over LTE or MPTCP is more energy-efficient than TCP over WiFi.
MPTCP with WiFi First can only utilize an LTE subflow when the WiFi subflow explicitly breaks, such as due to a WiFi AP disassociation.
For example, in our mobile scenario in Section \ref{subsec:exp-mbl}, MPTCP with WiFi First would not use the LTE subflow even when the WiFi subflow becomes unusable, e.g., the duration around 25$\sim$40 seconds in Figure \ref{fig:mgmt-comp-example-mobile}(a) and (b), since the device does not lose the WiFi association.
Therefore, if WiFi provides too low bandwidth to be more energy efficient than LTE while it is still associated, MPTCP with WiFi First degenerates into single-path TCP over WiFi, which yields inefficient energy usage as shown in the previous subsections.
In contrast, eMPTCP suspends and resumes an LTE subflow according to the estimated throughputs and energy efficiencies of each interface, regardless of the state of the association.
Thus, eMPTCP can adaptively control subflow usage so as to more quickly respond to available bandwidth changes and obtain more energy efficiency than MPTCP with WiFi First.

\section{Related Work}
\label{sec:related}



Balasubramanian et al.~\cite{Balasubramanian2009} measures energy
consumption on a Nokia N95 platform and identifies high energy overhead
as the result of the tail state in 3GPP interfaces (GSM, 3G).
They develop a protocol called {\em tailender} to schedule transfers
so as to minimize energy consumed by the tail.

Huang et al.~\cite{Huang2012close} provide an in-depth look at power and
performance characteristics of 4G LTE networks based on a large-scale
measurement of a commercial cellular provider.  They show that while
LTE is more energy-efficient than 3G, it is still not as efficient
as WiFi, partially due to the tail state cost in LTE.

Schulman et al.~\cite{schulman2010bartendr} show that the power consumed by
wireless radios is higher when the signal is week.  They present Bartendr,
a system for scheduling transmissions when the signal is strong so as to
minimize power consumption.
Ding et al.~\cite{Ding2013} provide a more in-depth analysis of the
impact of signal strength.

Rahmati et al. \cite{Rahmati2007context} also examine how to reduce power
consumption by utilizing the most efficient interface.  Noting that WiFi
is much more efficient than 3G, they devise algorithms to estimate WiFi
conditions without powering up the antenna, showing a 39\% improvement
via simulation in energy consumption.

Ra et al. \cite{Ra2010energy} show how a delay-tolerant application can
postpone its network usage until a more energy-efficient device is
available, saving 10$\sim$40\% of energy.

Bui et al. \cite{greenbag} implement a middleware to aggregate bandwidth of asymmetric wireless links for video streaming, called GreenBag. GreenBag estimates the available bandwidth of WiFi and LTE and determines the amount of allocated traffic to each interface while considering a quality-of-service requirement of streaming service. The authors show that GreenBag reduces energy consumption by 14$\sim$25\% compared with a bandwidth aggregation for throughput maximization. However,  since GreenBag operates above TCP layer as a system background process, it only works for modified HTTP requests sent not to original destinations but to GreenBag, therefore, each application needs to be modified to cooperate with GreenBag.

The most closely related work to ours is Pluntke et
al.~\cite{Pluntke2011} who introduce the concept of scheduling paths in MPTCP to minimize
energy consumption using a scheduler based on Markov decision processes.
The path scheduling decison is made periodically every ${t}$ time units.
Schedulers are computed in the cloud and downloaded periodically to the device.
They evaluate their scheduler via simulation, by using models of device
energy consumption and find they can reduce energy consumption by almost 10 \% in one
scenario (out of four), viewing high-quality video stream.
Our work, in contrast, is measured and evaluated experimentally using
a real MPTCP implementation on a physical device.

Raiciu et al.~\cite{Raiciu2011:Opportunistic} look at a number of issues
in using MPTCP for mobility, including power consumption.
They propose and evaluate a simple strategy that periodically samples both
paths for 10 seconds and then uses the more energy-efficient path
for 100 seconds.
Evaluating their strategy via simulation, the approach is more power
efficient than an energy-unaware MPTCP implementation, but achieves
lower bandwidth.

Paasch et al. \cite{Paasch2012} studied mobile/WiFi handover performance
with MPTCP.
The authors investigated the impact of handover on MPTCP connections using
different modes such as Full-MPTCP mode (where all potential subflows are
used to transmit packets) and Backup mode (where only a subset of subflows
are used).
They also measure energy consumption on a Nokia N950 smartphone for
two download scenarios and find that using WiFi alone is more
energy-efficient than base MPTCP.

\section{Conclusion}
\label{sec:conclusion}
This paper proposes a method to improve the energy efficiency of MPTCP. We develop eMPTCP, which manages subflows based on expected energy efficiency given available throughputs. We perform experiments with a real mobile device running MPTCP and eMPTCP. Our experimental results show that eMPTCP is able to consume less energy than MPTCP while it still provides the benefits of MPTCP such as robustness. Our results also show that eMPTCP obtains significantly better performance with comparatively small energy overhead or even with less energy consumption than TCP over WiFi. For future work, we will extend the implementation and experiments to reflect further environments, such as other mobile devices, as well as refining the parameter setting of eMPTCP to improve performance.

\section*{Acknowledgement}
This research was sponsored by US Army Research laboratory and
the UK Ministry of Defence and was accomplished under Agreement
No. W911NF-06-3-0001.
The views and conclusions contained
in this document are those of the authors and should not be interpreted
as representing the official policies, either expressed or implied,
of the US Army Research Laboratory, the U.S. Government, the
UK Ministry of Defense, or the UK Government. The US and UK
Governments are authorized to reproduce and distribute reprints for
Government purposes notwithstanding any copyright notation hereon.

\bibliographystyle{abbrv}
\bibliography{mptcp-power}

\end{document}